\documentclass[USenglish,oneside,twocolumn]{article}

\usepackage[utf8]{inputenc}
\usepackage[big]{dgruyter_NEW}
\usepackage[justification=centering]{caption}
\usepackage{amsfonts}
\usepackage{algorithm2e}
\usepackage{subcaption}
\usepackage{listings}
\usepackage{xspace}
\usepackage{threeparttable}
\usepackage{placeins}
\usepackage{float}
\usepackage{flushend}

\usepackage{amsmath}
\DeclareMathOperator*{\join}{\bowtie_\theta}

\newcommand{\eps}{\varepsilon}
\newcommand{\del}{\delta}
\newcommand{\epsdel}{(\eps,\del)}
\newcommand{\eDP}{$\eps$-DP\xspace}
\newcommand{\edDP}{$\epsdel$-DP\xspace}
\newcommand{\dbs}{\mathcal{D}}
\newcommand{\uid}{\ensuremath{\mathit{uid}}\xspace}
\newcommand{\me}{\mathrm{e}}

\newtheorem{definition}{Definition}
\newtheorem{dgthm}{Theorem}[]
\newtheorem{lemma}{Lemma}

\DOI{foobar}

\cclogo{\includegraphics{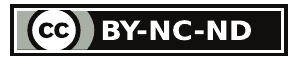}}
  
\begin{document}

\def\sectionautorefname{Section}
\def\subsectionautorefname{Section}
\def\subsubsectionautorefname{Section}
\def\paragraphautorefname{Section}
\def\definitionautorefname{Definition}
\def\algorithmautorefname{Algorithm}
\def\dgthmautorefname{Theorem}
\def\lemmaautorefname{Lemma}

\renewcommand\theparagraph{\thesubsection.\arabic{paragraph}}

  \author[1]{Royce J Wilson}
  \author[2]{Celia Yuxin Zhang}
  \author[3]{William Lam}
  \author[4]{Damien Desfontaines}
  \author[5]{Daniel Simmons-Marengo}
  \author[6]{Bryant Gipson}

  \affil[1]{Google, E-mail: rjwilson@google.com}
  \affil[2]{Google, E-mail: cyzhang@google.com}
  \affil[3]{Google, E-mail: lamw@google.com}
  \affil[4]{Google / ETH Zurich, E-mail: damien@desfontain.es}
  \affil[5]{Google, E-mail: dasm@google.com}
  \affil[6]{Google, E-mail: bryantgipson@google.com}

  \title{\huge Differentially Private SQL with Bounded User Contribution}

  \runningtitle{Differentially Private SQL}

  \begin{abstract}
  {Differential privacy (DP) provides formal guarantees that the output of a database query does not reveal too much information about any individual present in the database. While many differentially private algorithms have been proposed in the scientific literature, there are only a few end-to-end implementations of differentially private query engines. Crucially, existing systems assume that each individual is associated with at most one database record, which is unrealistic in practice. We propose a generic and scalable method to perform differentially private aggregations on databases, even when individuals can each be associated with arbitrarily many rows. We express this method as an operator in relational algebra, and implement it in an SQL engine. To validate this system, we test the utility of typical queries on industry benchmarks, and verify its correctness with a stochastic test framework we developed. We highlight the promises and pitfalls learned when deploying such a system in practice, and we publish its core components as open-source software.}
  \end{abstract}

  \journalname{Proceedings on Privacy Enhancing Technologies}
\DOI{Editor to enter DOI}
  \startpage{1}
  \received{..}
  \revised{..}
  \accepted{..}

  \journalyear{..}
  \journalvolume{..}
  \journalissue{..}

\maketitle

\section{Introduction} 

Many services collect sensitive data about individuals. These services must balance the possibilities offered by analyzing, sharing, or publishing this data with their responsibility to protect the privacy of the individuals present in their data. Releasing aggregate results about a population without revealing too much about individuals is a long-standing field of research. The standard definition used in this context is differential privacy (DP): it provides a formal guarantee on how much the output of an algorithm reveals about any individual in its input~\cite{dwork2014algorithmic, dwork2007ad, dwork2009differential}. Differential privacy states that the distribution of results derived from private data cannot reveal ``too much'' about a single person's contribution, or lack thereof, to that data~\cite{dwork2006calibrating}. By using differential privacy when analyzing data, organizations can minimize the disclosure risk of sensitive information about their users.

Query engines are a major analysis tool for data scientists, and one of the most common ways for analysts to write queries is with Structured Query Language (SQL). As a result, multiple query engines have been developed to enable data analysis while enforcing DP~\cite{mcsherry2009privacy, johnson2018towards, kotsogiannisarchitecting, bater2018shrinkwrap}, and all of them use a SQL-like syntax.

However, as we discuss in Section~\ref{sec:simple}, these differentially private query engines make some implicit assumptions, notably that each individual in the underlying database is associated with at most one database record. This does not hold in many real-world datasets, so the privacy guarantee offered by these systems is weaker than advertised for those databases. To overcome this limitation we introduce a generic mechanism for bounding user contribution to a large class of differentially private aggregate functions. We then propose a design for a SQL engine using these contribution bounding mechanisms to enforce DP, even when a given individual can be associated with arbitrarily many records or the query contains joins.

Our work goes beyond this design and accompanying analysis: we also describe the implementation of these mechanisms as part of a SQL engine, and the challenges encountered in the process. We describe a stochastic testing framework that generates databases on which we test for differential privacy to increase our level of trust into the system's robustness. To aid in replicability of our work and encourage wider adoption of differential privacy, we release core components of the system as open-source software.

\subsection{Requirements and contributions}

To be useful for non-expert analysis, a differentially private SQL engine must at least:

\begin{itemize}

\item Make realistic assumptions about the data, specifically allowing multiple records to be associated with an individual user.

\item Support typical data analysis operations, such as counts, sums, means, percentiles, etc.

\item Provide analysts with information about the accuracy of the queries returned by the engine, and give them clear privacy guarantees.

\item Provide a way to test the integrity of the engine and validate the engine's privacy claims.

\end{itemize}

In this work, we present a differentially private SQL engine that satisfies these  requirements. More precisely:

\begin{itemize}

\item We detail how we use the concept of \emph{row ownership} to enforce the original meaning of differential privacy: the output of the analysis does not reveal anything about a single \emph{individual}. In our engine, multiple rows can be associated with the same ``owner'' (hereafter referred to as a \emph{user}, although the owner could also be a group), and the differential privacy property is enforced at the user level.

\item We implement common aggregations (counts, sums, medians, etc.), arbitrary per-record transforms, and joins on the row owner column as part of our engine. To do so we provide a method of bounding query sensitivity and stability across transforms and joins, and a mechanism to enforce row ownership throughout the query transformation.

\item We detail some of the usability challenges that arise when trying to productionize such a system and increase its adoption. In particular, we explain how we communicate the accuracy impact of differential privacy to analysts, and we experimentally verify that the noise levels are acceptable in typical conditions. We also propose an algorithm for automatic sensitivity determination.

\item We present a testing framework to help verify that \eDP aggregation functions are correctly implemented, and can be used to detect software regressions that break the privacy guarantees.

\end{itemize}

Overall, this work contributes to research on differential privacy by proposing a method of bounded contributions and exploring the trade-offs in accuracy in a testable and verifiable way. Additionally, this work can increase the appropriate adoption of differential privacy by providing a usable system based on popular tools used by data analysts. For reproducibility and adoption, we release the new SQL aggregation operations and the stochastic tester as open-source software.

\subsection{Related work}

Multiple differentially private query engines have been proposed in the literature. In this work, we mainly compare our system to two existing differentially private query engines: PINQ~\cite{mcsherry2009privacy} and Flex~\cite{johnson2018towards}. Our work differs in two major ways from these engines: we support the common case where a single user is associated with multiple rows, and we support arbitrary \verb|GROUP BY| statements.

In these systems a single organization is assumed to hold all the raw data. Query engines can also be used in other contexts: differential privacy can be used in concert with secure multiparty computation techniques to enable join queries between databases held by different organizations, e.g., DJoin~\cite{narayan2012djoin} and Shrinkwrap~\cite{bater2018shrinkwrap}.

A significant amount of research focuses on improving the accuracy of query results while still maintaining differential privacy. In this work, for clarity, we keep the description of our system conceptually simple, and explicitly do not make use of techniques like smooth sensitivity~\cite{nissim2007smooth}, tight privacy budget computation methods~\cite{kairouz2017composition,meiser2018tight}, variants of the differential privacy definition~\cite{bun2016concentrated,mironov2017renyi,desfontaines2019sok}, adjustment of noise levels to a pre-specified set of queries~\cite{li2014data}, or generation of differentially private synthetic data to answer arbitrarily many queries afterwards~\cite{bindschaedler2017plausible,kotsogiannisarchitecting,kotsogiannis2019privatesql}.

The testing framework we introduce in~\autoref{sec:stochastictesting} is similar to recent work in verification for differential privacy~\cite{DingEtAl2018,bichsel2018dp}, but approaches the problem in a generic way by testing a diverse set of databases that are agnostic to specific algorithms.

Our work is not the first to use noise and thresholding to preserve privacy: this method was originally proposed in~\cite{korolova2009releasing,gotz2011publishing} in the specific context of releasing search logs with \edDP; our work can be seen as an extension and generalization of this insight. Diffix~\cite{francis2017diffix} is another system using similar primitives; however, it does not provide any formal privacy guarantee; so a meaningful comparison with our work is not feasible. In \autoref{sec:experiments}, we provide a comparison of query accuracy between our work, PINQ, and Flex.

\subsection{Preliminaries}\label{sec:preliminaries}

We introduce here the definitions and notations used throughout this paper. Let $R$ be an arbitrary set of \emph{records}, and $\mathcal{U}$ an arbitrary set of \emph{user identifiers}. A \emph{row} is a pair $(u,r)$ for some $u\in \mathcal{U}$ and $r\in R$, and a \emph{database} is a multiset of rows. A user $u$ is said to \emph{own} the rows $(u,r)$ for all $r\in R$. We denote $\dbs$ the space of all databases. 

\begin{definition}[Distance between databases]
We denote \emph{row-level change} the addition or removal of a single row from a database, and \emph{user-level change} the addition or removal of all rows associated with a user. Given two databases $D_1$ and $D_2$, we denote $||D_1-D_2||$ the minimum number of row-level changes necessary to transform $D_1$ into $D_2$, and ${||D_1-D_2||}_u$ the minimum number of user-level changes necessary to transform~$D_1$ into~$D_2$.
\end{definition}

We recall the definitions of $(\eps,\del)$-differential privacy and of global $L^1$-sensitivity.

\begin{definition}[$(\eps,\del)$-Differential Privacy]\label{differentialprivacy}
  A randomized mechanism $f\colon \dbs\rightarrow \mathbb{R}^d$ satisfies \emph{row-level \edDP} if for all pairs of databases $D_1,D_2\in\dbs$ that satisfy $||D_1-D_2||=1$, and for all sets of outputs $S$, we have:
  \[
  \Pr[f(D_1) \in S] \leq  \me^{\eps} \; \Pr[f(D_2) \in S] + \del
  \]
  $f$ satisfies \emph{user-level DP}\footnote{A similar notion in the context of streaming data, \emph{pan-privacy}, is introduced in~\cite{dwork2010pan}.} if the above condition holds for all pairs of databases $D_1,D_2\in\dbs$ such that ${||D_1-D_2||}_u=1$. \eDP is an alias for $(\eps,0)$-DP\xspace.
\end{definition}

Note that this notion is technically \emph{unbounded differential privacy}~\cite{kifer2011no}, which we use for simplicity throughout this work. Up to a change in parameters, it is equivalent to the classical definition, which also allows the \emph{change} of one record in the distance relation between databases.

\begin{definition}[$L^1$-Sensitivity]\label{def:sensitivity}
  The \emph{global $L^1$-sensitivity} of a function $f\colon \dbs\rightarrow \mathbb{R}^d$ is defined by:
    \[\Delta f = \underset{D_1,D_2\in\dbs:||D_1-D_2||=1}\max{{||f(D_1)-f(D_2)||}_1}
    \]
  where $||.||_1$ denotes the $L^1$ norm. The \emph{user-global $L^1$-sensitivity} of $f$ is defined by:
    \[
    \Delta_u f = \underset{D_1,D_2\in\dbs:{||D_1-D_2||}_u=1}\max{{||f(D_1)-f(D_2)||}_1}
    \]
\end{definition}

\section{A simple example: histograms}\label{sec:simple}

Before describing the technical details of our system we first give an intuition of how it works using a simple example: histogram queries. Consider a simple database that logs accesses to a given website. An analyst wants to know which browser agents are most commonly used among users visiting the page. A typical query to do so is presented in~\autoref{listing:simple}.

\begin{lstlisting}[caption={Simple histogram query},captionpos=b,label={listing:simple}]
SELECT browser_agent, COUNT(*) AS visits
FROM access_logs
GROUP BY browser_agent;
\end{lstlisting}

How would one make this simple operation $\eps$-differentially private? One naive approach is to add Laplace noise of scale $1/\eps$ to each count. This solution suffers from several shortcomings.

\subsection{First pitfall: multiple contributions within a partition}

The naive approach will correctly hide the existence of individual \emph{records} from the database: each record of the access log will only influence one of the returned counts by at most $1$, and it is well known~\cite{dwork2006calibrating} that this mechanism will provide \eDP. However, it fails to protect the existence of individual \emph{users}: the same user could have visited the example page many times with a particular browser agent, and therefore could have contributed an arbitrarily large number of rows to \verb|visits| for a particular \verb|GROUP BY| partition, violating our assumption that query sensitivity is 1.

In PINQ and Flex, the differential privacy definition explicitly considers records as the privacy unit. Because we instead want to protect the full contribution of users, we need to explicitly include a notion of a user in our system design. In this work, we do this via the notion of a user identifier, hereafter abbreviated \emph{uid}.

Because \autoref{listing:simple} has unbounded sensitivity, adding noise to counts is not enough to enforce differential privacy; we need to \emph{bound user contribution to each partition}. This can be addressed is by counting distinct users, which has a user-global sensitivity of 1, instead of counting rows. Although this modifies query semantics, we chose this approach to keep the example simple. We present the modified query in~\autoref{listing:fix1}.

\begin{lstlisting}[caption={Partition level contribution bounding},captionpos=b, label={listing:fix1}]
SELECT browser_agent,
       COUNT(DISTINCT uid) + Laplace(1/$\eps$)
FROM access_logs
GROUP BY browser_agent;
\end{lstlisting}

In other contexts it might make more sense to allow a user to contribute more than once to a partition (e.g. we count up to five visits from each distinct user with each distinct browser agent); in this case we would need to further modify the query to allow multiple contributions and increase sensitivity to match the maximum number of contributions.

\subsection{Second pitfall: leaking GROUP BY keys}\label{sec:leakingkeys}

Even if we bound contribution to partitions and adapt noise levels, the query is still not \eDP. Suppose that the attacker is trying to distinguish between two databases differing in only one record, but this record is a unique browser agent ${\emph{BA}}_{unique}$: this browser agent does not appear in $D_1$, but appears once in $D_2$. Then, irrespective of the value of the noisy counts, the \verb|GROUP BY| keys are enough to distinguish between the two databases simply by looking at the output: ${\emph{BA}}_{unique}$ will appear in the query output for $D_2$ but not for $D_1$.

A simple solution to this problem was proposed in \cite{korolova2009releasing}: the idea is to drop from the results all keys associated with a noisy count lower than a certain threshold $\tau$. $\tau$ is chosen independently of the data, and the resulting process is \edDP with $\del>0$. We call this mechanism \emph{$\tau$-thresholding}. With a sufficiently high $\tau$, the output rows with keys present in $D_2$ but not $D_1$ (and vice-versa) will be dropped with high probability, making the keys indistinguishable to an attacker. A longer discussion on the relation between $\eps$, $\del$ and $\tau$ can be found in \autoref{sec:kthreshold}. This approach is represented in SQL in~\autoref{listing:fix2}.

\begin{lstlisting}[caption={GROUP BY filtering},captionpos=b, label={listing:fix2}]
SELECT browser_agent,
       COUNT(DISTINCT uid) + Laplace(1/$\eps$) AS c
FROM access_logs
GROUP BY browser_agent
HAVING c >= $\tau$;
\end{lstlisting}

PINQ and Flex handle this issue by requiring the analyst to enumerate all keys to use in a \verb|GROUP BY| operation, and return noisy counts for only and all such keys\footnote{The open-source implementation of Flex~\cite{flexgithub}, however, does not appear to implement this requirement.}. This enforces \eDP but impairs usability: the range of possible values is often large (potentially the set of all strings) and difficult to enumerate, especially if the analyst cannot look at the raw data.

Some data synthesis algorithms have been proposed to release histograms under \eDP~\cite{mcsherrydatasynthesis}, but are limited, for example to datasets subject to hierarchical decomposition. Our approach is simpler and more generic, at some cost in the privacy guarantee.

\subsection{Third pitfall: contributions to multiple partitions}\label{sec:pitfallsampling}

Finally, we must consider the possibility of a user contributing to multiple partitions in our query. Imagine a user visiting the example page with many different browsers, each with a different browser agent. Such a user could potentially contribute a value of 1 to each partition's count, changing the sensitivity of the query to be the number of partitions, which is unbounded!

Because both PINQ and Flex consider records as the privacy unit, this is not an issue for their privacy models. So long as they are only used on databases where that requirement holds true, and where the sensitivity and stability impact of joins (and related operations) are carefully considered, they will provide adequate DP guarantees. However as shown in~\cite{ubernotdp}, these conditions are not always true.

Instead of adding strict requirements on the nature of the underlying database and on how joins are used, we introduce a novel mechanism for \emph{bounding user contribution across partitions}. Concretely, we first choose a number $C_u$, and for each user, we randomly keep the contributions to $C_u$ partitions for this user, dropping contributions to other partitions. This operation allows us to bound the global sensitivity of the aggregation: each user can then influence at most unique $C_u$ counts, and we can adapt the noise level added to each count, by using Laplace noise of scale $C_u/\eps$.

The final version of our query is shown in~\autoref{listing:fix3}. It uses a non-standard variant of the SQL \verb|TABLESAMPLE| operator, which supports partitioning and reservoir sampling, to represent the mechanism we introduced. This final version satisfies $(\eps,\del)$-differential privacy for well-chosen parameters.

\begin{lstlisting}[caption={An \edDP query},captionpos=b, label={listing:fix3}]
SELECT browser_agent,
       COUNT(DISTINCT uid) + Laplace($C_u$/$\eps$) AS c
FROM (SELECT browser_agent, uid
      FROM access_logs
      GROUP BY browser_agent, uid)
TABLESAMPLE RESERVOIR
  ($C_u$ ROWS PARTITION BY uid)
GROUP BY browser_agent
HAVING c >= $\tau$;
\end{lstlisting}

In the remainder of this paper, we formalize this approach, and adapt it to a larger set of operations. In particular, we extend it to arbitrary aggregations with bounded sensitivity, and we explain how to make this model compatible with joins.

\section{System model and design} 

\subsection{Overview}\label{sec:overview} 

As suggested in Section~\ref{sec:preliminaries}, we assume that there is a special column of the input database that specifies which user owns each row. The system is agnostic to the semantics of this special column. In principle, it can be any unit of privacy that we need to protect: a device identifier, an organization, or even a unique row ID if we want to protect rows and not users. For simplicity of notation we assume that this special column is a user identifier. Users may own multiple rows in each input table, and each row must be owned by exactly one user. Our model guarantees \edDP with respect to each user, as defined in \autoref{differentialprivacy}.

We implement our DP query engine in two components on top of a general SQL engine: a collection of custom SQL aggregation operators (described in \autoref{sec:boundedagg}), and a query rewriter that performs anonymization semantics validation and enforcement (described in \autoref{sec:querysemantics}). The underlying SQL engine tracks user ID metadata across tables, and invokes the DP query rewriter when our anonymization query syntax is used on tables containing user data and any applicable permission checks succeed. \autoref{listing:example} provides an example of a SQL query accepted by our system.

\begin{lstlisting}[caption={Anonymization query example},captionpos=b, label={listing:example}]
SELECT WITH ANONYMIZATION
  T1.cohort, ANON_SUM(T2.val, 0, 1)
FROM Table1 T1, Table2 T2 USING(uid)
GROUP BY T1.cohort;
\end{lstlisting}

The query rewriter decomposes such queries into two steps, one before and one after our introduced DP aggregation operator, denoted by \verb|SELECT| \verb|WITH| \verb|ANONYMIZATION|. The first step begins by validating that all table subqueries inside the DP operator's \verb|FROM| clause enforce unique user ownership of intermediate rows. Next, for each row in the subquery result relation, our operator partitions all input rows by the vector of user-specified \verb|GROUP BY| keys and the user identifier, and applies an intermediate vanilla-SQL partial aggregation to each group. 

For the second step, we sample a fixed number of these partially aggregated rows for each user to limit user contribution \emph{across partitions}. Finally, we compute a cross-user DP aggregation across all users contributing to each \verb|GROUP BY| partition, limiting user contribution \emph{within partitions}. Adjusting query semantics is necessary to ensure that, for each partition, the cross-user aggregations receive only one input row per user.

\subsection{Bounded-contribution aggregation}\label{sec:boundedagg} 

In this section, we present the set of supported \eDP statistical aggregates with bounded contribution. These functions are applied as part of the cross-user aggregation step, discussed in \autoref{sec:querysemantics}. Importantly, at this step, we assume that each user's contributions have been aggregated to a single input row - this property is enforced by the query rewriter.

For a simple example for bounded contribution, \verb|COUNT(DISTINCT uid)| counts unique users. Adding or subtracting a user will change the count by no more than 1.

For more complex aggregation functions we must determine how much a user can contribute to the result and add appropriately scaled noise. A naive solution without limits on the value of each row leads to unbounded contribution by a single user. For example, a \verb|SUM| which can take any real as input has an unbounded $L^1$-sensitivity by~\autoref{def:sensitivity}.

To address this, each \eDP function accepts an additional pair of lower and upper limit parameters used to clamp (i.e., \emph{bound}) each input. For example, denoting the lower and upper bounds as $L$ and $U$, respectively, consider the anonymized sum function:
\[
\verb|ANON_SUM(col, L, U)|
\]

Let $\text{sum}_L^U\colon \dbs\rightarrow\mathbb{R}$ be the function that transforms each of its inputs $x$ into $x'=\max(\min(x,U),L)$, and then all $x'$ are summed. The global sensitivity for this bounded sum function is:
  \[
  \Delta \text{sum}_L^U = \max (|L|, |U|)
  \]
and thus, \verb|ANON_SUM| can be defined by using this function and then adding noise scaled by this sensitivity. For all functions, noise is added to internal states using the well-known Laplace mechanism~\cite{geng2012optimal,dwork2014algorithmic} before a differentially private version of the aggregate result can be released.
 
For \verb|ANON_AVG|, we use the algorithm designed in Li et al.~\cite{li2016differential}: we take the quotient of a noisy sum (bounded as in \verb|ANON_SUM|, and scaled by sensitivity $|U-L|/2$) and a noisy count. \verb|ANON_VAR| is similarly derived; we use the same algorithm to compute a bounded mean and square it, and to compute a mean of bounded squares. \verb|ANON_STDDEV| is implemented as the square root of \verb|ANON_VAR|.

Lastly, \verb|ANON_NTILE|, is based on a Bayesian binary search algorithm~\cite{ben2008bayesian, karp2007noisy}, and can be used to define max, min, and median functions. The upper and lower bounds are only used to restrict the search space and do not affect sensitivity. Each iteration of the internal binary search alters counts with a noise-adding mechanism scaled by sensitivity $1$. 

For the rest of this paper, we assume that contribution bounds are specified as literals in each query to simplify our presentation. Setting bounds requires some prior knowledge about the input set. For instance, to average a column of ages, the lower bound could be reasonably set to $0$ and the upper bound to $120$. To enhance usability in the case where there is no such prior knowledge, we also introduce a mechanism for automatically inferring contribution bounds, described in more detail in~\autoref{approxbounds}.

The definition of sensitivity given in~\autoref{def:sensitivity} assumes deterministic functions. From here, we will say that the \emph{global sensitivity of a \eDP aggregate function is bounded by $M$} if the global sensitivity of the same function with no noise added is bounded by $M$. Due to the contribution bounding discussed in this section, we can determine such bounds $M$ for our \eDP aggregation functions.

\autoref{table:agg} lists our suite of aggregate functions and their sensitivity bounds, proven in~\autoref{app:sensitivitybound}. These bounds assume that the aggregation is done on at least one user; we do not consider the case where we compare an empty aggregation with an aggregation over one user. This last case is considered in \autoref{sec:kthreshold}. Note that the bounds shown here are loose; we mostly care about the boundedness, and the order of magnitude with respect to $U$ and $L$.

\begin{table}
  \caption{\eDP aggregate functions}\label{table:agg}
	\centering
	\begin{tabular}{ll}
		\toprule
		Function & Sensitivity Bound\\
		\midrule
		\verb|ANON_COUNT(col)| & $1$ \\
		\verb|ANON_SUM(col, L, U)| & $\max(|L|, |U|)$ \\
		
		\verb|ANON_AVG(col, L, U)| & $|U-L|$\\
		\verb|ANON_VAR(col, L, U)| & $ |U-L|^2$  \\
		\verb|ANON_STDDEV(col, L, U)|& $|U-L|$ \\
		
		\verb|ANON_NTILE(col, ntile, L, U)| & $|U-L|$\\
		\bottomrule
	\end{tabular}
\end{table}

\subsection{Query semantics}\label{sec:querysemantics} 

In this section, we define our \edDP relational operator, denoted by $\chi$. To do so, we use some of the conventional operators in relational algebra, a language used to describe the transformations on databases that occur in a query:

\begin{itemize}
    \item $\Pi_s(R)$: Project columns $s$ from $R$.
    \item $\sigma_\varphi(R)$: Select from $R$ satisfying the predicate $\varphi$.
    \item $ _{g}\mathcal{G}_{a}(R)$: Group on the cross products of distinct keys in the columns in $g$. Apply the aggregations in $a$ to each group.
    \item $R \join\limits_{\text{cond}} S$: Take the cross product of rows in $R$ and $S$, select only the rows that satisfy the predicate $\text{cond}$.
\end{itemize}

Let $s$ (select-list), $g$ (group-list), and $a$ (aggregate-list) denote the attribute names $s_1, \dots ,s_i$, $g_1, \dots ,g_j$, and $a_1, \dots ,a_k$, respectively. Assume $a$ is restricted to only contain the \eDP aggregate function calls discussed in \autoref{sec:boundedagg}. Our introduced operator, $_{g}\chi_{a}$, can be interpreted as an anonymized grouping and aggregation operator with similar semantics to $_{g}\mathcal{G}_{a}$.

Let $T$ be a table subquery containing any additional analyst-specified operators that do not create any intermediate objects of shared ownership (as defined in \autoref{peruserpartitioning}). Let $R$ be an input table with each row owned by exactly one user. $R$ must have a denoted user-identifying \emph{uid} attribute in the schema for the query to be allowed. We use $T(R)$ as the input to our proposed operator $\chi$. Then, we can define our query $Q$:
\[
Q := \Pi_{s} (_{g}\chi_{a} (T(R)))
\]

Our approach represents the general form of this relational expression using augmented SQL:
\begin{lstlisting}
SELECT WITH ANONYMIZATION $s$, $a$
FROM $T$(R)
GROUP BY $g$;
\end{lstlisting}

The query rewriter discussed in \autoref{sec:overview} transforms a query containing $\chi$ into a query that only contains SQL primitives, minimizing the number of changes to the underlying SQL engine. We define the following in order to express our rewriter operation:

\begin{itemize}
\item Let \uid be the unique user identifier associated with each row. Let $f(\uid)$ compute the additional privacy risk for releasing a result record. $f$ represents the \emph{$\tau$-thresholding} mechanism introduced in \autoref{sec:leakingkeys} and its nature is discussed in more detail in \autoref{sec:kthreshold}.

\item Let $a'$ be the corresponding non-\eDP partial aggregation function of $a$. For example, if $a$ is \verb|ANON_SUM|, $a'$ would be \verb|SUM|. 

\item Let $_m\mathcal{T}_n$ behave like a reservoir-sampling SQL \verb|TABLESAMPLE| operator where $m$ is the grouping list and $n$ is the number of samples per group. In other words, the operation  $_m\mathcal{T}_n(R)$ partitions $R$ by columns $m$. For each partition, it randomly samples up to $n$ rows. We use $\mathcal{T}$ as the \emph{stability-bounding operator} and discuss its implications in \autoref{querystability}.
\end{itemize}

When the query rewriter is invoked on the following relational expression containing $\chi$:
\[
Q := \Pi_{s} (_{g}\chi_{a} (T(R)))
\]
it returns the modified expression, with $\chi$ expanded:
\begin{align*}
U &:= \Pi_{\uid, g, a'}(_\uid\mathcal{T}_{C_u}(_{\uid, g}\mathcal{G}_{a'}(T(R)))\\
S &:= \Pi_{s}\sigma_{f(\uid) < \delta}(_{g}\mathcal{G}_{f(\uid), a} (U))
\end{align*}

Effectively, the rewriter splits $Q$ into the two-stage aggregation $U$ and $S$. $U$ groups the output of $T(R)$ by the key vector $(\uid,g)$, applying the partial aggregation functions $a'$ to each group. $C_u$ rows are then sampled for each user. The first aggregation enforces that there is only one row per user per partition during the next aggregation step. $S$ performs a second, differentially private, aggregation over the output of $U$. This aggregation groups only by the keys in $g$ and applies the \eDP aggregation functions $(f(\uid),a)$. These functions assume that every user contributes at most one input row. A filter operator is applied last to suppress any rows with too few contributing users.

\vspace{-0.5cm}
\paragraph{Allowed subqueries} 
\label{peruserpartitioning}

In this section, we introduce the constraints imposed by $\chi$ on the table subquery $T$. Our approach requires that the relational operators composing $T$ do not create any intermediate objects of shared ownership, that is, no intermediate row may be derived from rows owned by different users. A naive application of certain relational operators violate this requirement. For example, for the aggregation operator, rows owned by distinct users may be aggregated into the same partition. Then the resulting row from that partition will be owned by all users whose data are contributed to the group. For the naive join operator, two rows from distinct users may be joined together, creating a row owned by both users. 

This restriction limits our system since some queries cannot be run. We observed that in practice, most use-cases can be fit within these constraints. We leave extensions to a wider class of queries for future work. This might require a different model than the one presented here, but changing query semantics will always be necessary for queries with unbounded sensitivity.

We address the shared ownership issue by restricting each operator composing $T$ such that, for each row in that operator's output relation, that row is derived only from rows in the input relation that have matching \emph{uid} attributes. We enforce this rule for aggregate operators by requiring that the analyst additionally groups-by the input relation's \emph{uid} attribute. For join operators, we require that the analyst adds a \verb|USING|(\uid) clause (or equivalent) to the join condition. Additionally, each operator of $T$ must propagate \emph{uid} from the input relation(s) to the output relation. In queries where this is unambiguous (i.e., the analyst does not refer to \emph{uid} in the query), we can automatically propagate \emph{uid}. 

Allowed alternatives for each basic operator are listed in \autoref{rewrite-table}. They are enforced recursively during the query rewrite for each operator that composes~$T$. 

\begin{table}
    \caption{Allowed Table Subquery Operators}
	\label{rewrite-table}
	\centering
	\begin{tabular}{lll}
		\toprule
		Operator & Basic Form & Required Variant \\
		\midrule
		Projection & $\Pi_{a}(R)$ &  $\Pi_{\uid, a}(R')$ \\
		Selection & $\sigma_{\varphi}(R)$ &  $\sigma_{\varphi}(R')$ \\
		Aggregation & $_{g}\mathcal{G}_{a}(R)$ & $_{\uid, g}\mathcal{G}_{a}(R')$ \\
		Join & $R \join\limits_{\text{cond}} S$ & $\underset{\text{cond} \wedge r.\uid=s.\uid}{R' \hfill \join \hfill S'}$  \\

		\bottomrule
	\end{tabular}
\end{table}

\vspace{-0.5cm}
\paragraph{Example: two-step aggregation}

Consider the following query, counting the number of employees per department with at least 
one order:
\begin{lstlisting}
SELECT WITH ANONYMIZATION
  dept, ANON_COUNT(*, 0, 5) as c
FROM Employee E, Order O USING(uid)
GROUP BY dept;
\end{lstlisting}

Note that including bounds on \verb|ANON_COUNT(*, L, U)| is shorthand for using \verb|ANON_SUM(col, L, U)| in the cross-user aggregation step. We can express this query in relational algebra using our DP operator, $\chi$:
\begin{align*}
Q := \; &\Pi_{\text{dept}, c}(_{\text{dept}} \chi _{\text{anon\_count}(*, 0, 5)\; \text{as}\; c }(\text{E} \join\limits_{\text{E.\uid} = \text{O.\uid}} \text{O}))
\end{align*}

Expand $\chi$ into the two-stage aggregation, $U$ and $S$. $T$ is a table subquery. Our query can then be written as:
\begin{align*}
T := \; &\Pi_{\text{dept}, \uid, c_2} (_{\text{dept}, \uid, } \mathcal{G}_{\text{count}(*) \; \text{as} \; c_2} (\text{E} \join\limits_{\text{E.\uid} = \text{O.\uid}} \text{O})) \\
U := \; &\Pi_{\text{dept}, uid, c_2}
      (_\uid\mathcal{T}_{C_u}
      (U)) \\
S  := \; &\Pi_{\text{dept}, c} \;\sigma_{c_3 \geq \tau}(_{\text{dept}}\mathcal{G}
\begin{aligned}[t]
_(&_{\text{anon\_sum}(c', 0, 5)\; \text{as}\; c, \;} \\
&_{\text{anon\_count}(*) \; \text{as} \; c_3)}(T))
\end{aligned}
\end{align*}

Note that we add an additional user counting \eDP function, aliased as $c_3$, which is compared to our threshold parameter, $\tau$, to ensure that the grouping does not violate the  $\epsdel$-DP predicate, to be discussed in \autoref{sec:kthreshold}. In this case $c_3$ is the number of unique users in each department. 

\begin{figure*}[th!]
    \centering
    \includegraphics[width=0.85\textwidth]{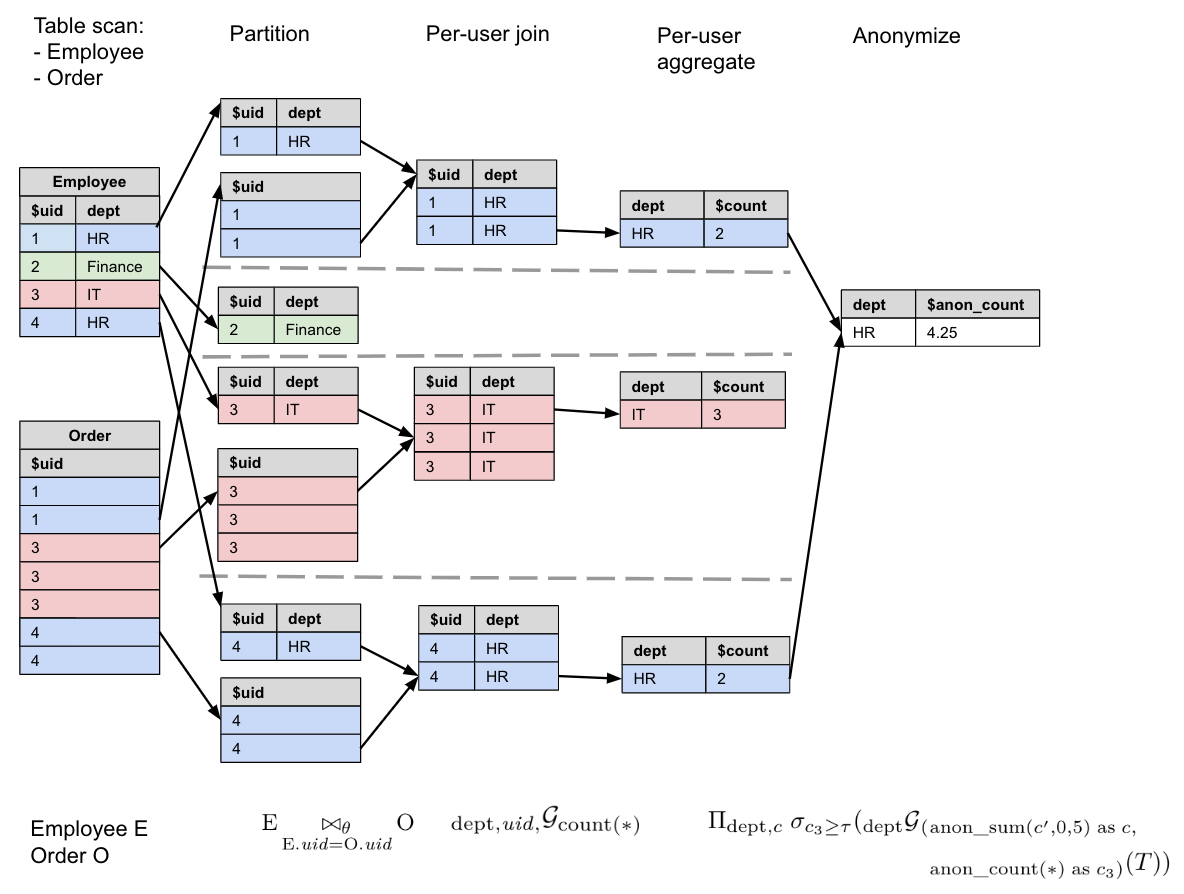}
    \caption{Example workflow of anonymized query. Note that in the last step, the IT department gets dropped from $\tau$-thresholding.}
    \label{fig:twostep}
\end{figure*}

In \autoref{fig:twostep}, we illustrate the workflow with example tables \verb|Employee| and \verb|Order|, $\tau=2$, and $C_u=1$. 

\subsection{Query stability and sensitivity}\label{querystability} 

\vspace{-0.2cm}
\paragraph{Bounding stability}

We adapt the notion of query stability from~\cite{mcsherry2009privacy}.

\begin{definition}[Global Stability]\label{stability}
  Let $T$ be a function $T\colon \dbs \rightarrow \dbs$. We say that $T$ has \emph{$c$-stability} if for all $D_1,D_2\in\dbs$:
  \[
  ||T(D_1) - T(D_2)|| \leq c||D_1-D_2||.
  \]
\end{definition}

Note that for a user $u$ owning $k$ rows in $D$ and a $c$-stable transformation $T$, there may be $k \cdot c$ rows derived from rows owned by $u$ in $T(D)$.

Our privacy model requires the input to the cross-user aggregation to have constant stability. Simple SQL operators have a stability of one. For instance, each record in an input relation can only be projected onto one record. So an addition or a deletion can only affect one record in a projection; thus projections have a stability of one. The same logic applies for selection. Other operators, such as joins, have unbounded stability because source records can be multiplied. Adding or removing a source record can affect an unbounded number of output records. When SQL operators are sequentially composed, we multiply the stability of each operator to yield the entire query's overall stability.

We can compose an unbounded transform $T$ with a stability-bounding transform $\mathcal{T}_{C_u}$ to yield a composite $C_u$-stable transform: 
\[
\left||\mathcal{T}_{C_u}(T(D_1)) - \mathcal{T}_{C_u}(T(D_2))|\right| \leq {C_u} \left||D_1 - D_2|\right|
\]

For $\mathcal{T}_{C_u}$, we use partitioned-by-user reservoir sampling with a per-partition reservoir size of $C_u$, which has a stability of $C_u$. Reservoir sampling was chosen for its simplicity, and because it guarantees a strict bound on the contribution of one user. Non-random sampling (e.g. taking the first $C_u$ elements, or using systematic sampling) risks introducing bias in the data, depending on the relative position of records in the database. Simple random sampling does not guarantee contribution bounding, and all types of sampling with replacement can also introduce bias in the data.

Joins appear frequently in queries~\cite{johnson2018towards}, so it is imperative to support them in order for an engine to be practical. Since joins have unbounded stability, a stability bounding mechanism is necessary to provide global stability privacy guarantees on queries with joins. We can thus support a well bounded, full range of join operators. 

\vspace{-0.5cm}
\paragraph{Bounding sensitivity}

In this section, we show that the user-global sensitivity of any allowed query in our engine is bounded. Sensitivity is bounded due to the structure of our two-stage aggregation, the bounded-contribution aggregation functions, and the stability-bounding operator $\mathcal{T}$.

\begin{dgthm}\label{sensitivitytheorem}
Consider an anonymization query in the form $h(R)$, where $R$ is the input table and $h$ is the query transformation. Then there exist constants $M$, which depends on $h$, and an engine-defined constant $C_u$, such that the user sensitivity satisfies  $\Delta_u h \leq C_u M$.
\end{dgthm}

\begin{proof} 
\autoref{app:sensitivityproof}.
\end{proof}

\subsection{Minimum user threshold}\label{sec:kthreshold} 

In this section, we outline a technique proposed in~\cite{korolova2009releasing} to prevent the presence of arbitrary group keys in a query from violating the privacy predicate: the \emph{$\tau$-thresholding} mechanism. For example, consider the naive implementation in~\autoref{listing:threshold}:

\begin{lstlisting}[caption={Leaking GROUP BY keys},captionpos=b,label={listing:threshold}]
SELECT col1, ANON_SUM(col2, L, U) 
FROM Table
GROUP BY col1;
\end{lstlisting}

Suppose that for the value $\verb|col1| = c$, only one user $u$ contributed to the sum. Then querying without data from $u$ would reveal the absence of the output row corresponding to group $\verb|col1| = c$. It would be revealed with certainty that user $u$ has value $c$ for $\verb|col1|$.

To prevent this, for each grouping or aggregation result row we first calculate an \eDP count of unique contributing users. If that count is less than some \emph{minimum user threshold} $\tau$ the result row must not be released. $\tau$ is chosen by our model based on $\eps$, $\delta$, and $C_u$ parameter values. In the example above, the output row for group $\verb|col1| = c$ would not appear in the result with some probability.

\begin{dgthm}\label{deltatheorem}
  Let $\eps, \delta, C_u > 0$ be privacy parameters. Consider a SQL engine that, for each non-empty group in a query's grouping list, computes and releases an $\eps/C_u$-DP noisy count of the number of contributing users. For empty groups, nothing is released. A user may not influence more than $C_u$ such counts. Each count must be $\tau$ or greater in order to be released. We may set
    \[
    \tau = 1 - \frac{C_u \log(2-2(1-\delta)^{1/C_u})}{\eps}
    \]
  to provide user-level $(\eps, \del)$-DP in such an engine.
\end{dgthm}

The proof to ~\autoref{deltatheorem} is supplied in~\autoref{app:deltaproof}. Our engine applies $\tau$-thresholding after the per-user aggregation step. Thus, we can generalize \autoref{deltatheorem} to our engine by using composition theorems for differential privacy.

\subsection{User-level differential privacy}\label{peruserprivacy} 

In this section, we show that our engine satisfies user-level \edDP, as defined in \autoref{differentialprivacy}. 

Suppose that a query $f$ has aggregate function list $a = \{a_1, \dots, a_N\}$ and grouping list $g = \{g_1, \dots, g_J\}$. Let the privacy parameter for aggregation function $a_i$ be $\eps_i$. 

Consider any set of rows owned by a single user in the input relation of our anonymization operator, $_g \chi_a$. We first partition and aggregate these rows by the key vector $(\text{uid},g)$, before sampling up to $C_u$ rows for each partition by $(\text{uid})$.

The result for a group $j$ is reported if the $\tau$-thresholding predicate defined in \autoref{sec:kthreshold} is satisfied.
Computing and reporting the comparison count for this predicate is $\eps_j'$-DP. For $D_1,D_2 \in \dbs$, such that they differ by a user's data for a \emph{single group} $j$, consider each aggregation function $a_i$ as applied to group $j$. By composition theorems~\cite{kairouz2017composition} and \autoref{deltatheorem}, we provide $(\eps_j' + \sum \eps_i, \del_j)$ user-level \edDP for that row.

However, there are many groups in a given query. Due to our stability bounding mechanism, a single user can contribute to up to $C_u$ groups. The $C_u$ output rows corresponding to these groups can be thought of as the result of $C_u$ sequentially composed queries. Let $D_1, D_2 \in \dbs$ such that $||D_1-D_2||_u=1$. Let $\eps$ be the sum of the $C_u$ greatest elements in the set $\{\eps_j' + \sum \eps_i\}_{j = 1, \dots, J}$. By composition theorems in differential privacy~\cite{kairouz2017composition}, we conclude that for any output set $S$, and some $\delta > 0$, we have  
\[
\Pr[f(D_1) \in S] \leq e^{\eps} \Pr[f(D_2) \in S] + \delta
\]

This shows that given engine-defined parameters $\eps$, $\delta$, and $C_u$, it is possible to set privacy parameters for individual \eDP functions to satisfy the query \edDP predicate. In reality, we conservatively set $\eps_j', \eps_i = \frac{\eps}{C_u (N+1)}$ for all $i$ and $j$. $\delta$ is enforced by the derived parameter $\tau$ as discussed in \autoref{sec:kthreshold}. Both user-privacy parameters $\eps$ and $\delta$ can therefore be bounded to be arbitrarily small by analysts and data owners. 

Note that the method we use to guarantee user-level differential privacy can be interpreted as similar to row-level group privacy: after the per-user aggregation step, we use the composition theorem to provide group privacy for a group of size $C_u$. Alone, row-level group privacy does not provide user-level privacy, but in combination to user-level contribution bounding, this property is sufficient to obtain the desired property.

\section{Accuracy}\label{sec:experiments}

In this section, we explore the accuracy of our system by running numerical experiments and provide analytical reasoning about the relationship between accuracy and various parameters.

\subsection{Experimental accuracy}

\begin{table*}
	\centering
\begin{threeparttable}
    \caption{TPC-H Query 1 errors comparison with others ($\eps=0.1$)}
	\label{comparison-table}

	\begin{tabular}{lllll}
		\toprule
		Function & $\Delta_u$Q1 & Our model & PINQ & Flex \\
		\midrule
		$\eps$-COUNT(*) & $373$ & $0.00175$ &  $0.00179$ & $0.0197$ \\
		$\eps$-AVG(l\_extendedprice) & $100000$ & $0.00181$ & $0.00181$ & \tnote{\textasteriskcentered} \\
		$\eps$-MEDIAN(l\_extendedprice) & $100000$ & $0.00189$ & $0.00349$ & \tnote{\textasteriskcentered} \\
		$\eps$-COUNT(*) & $1$\tnote{$\dagger$} & $0.993$ & $4.727\times10^{-6}$ & $2.70\times 10^{-5}$  \\
		\bottomrule
	\end{tabular}
	\begin{small}
	\begin{tablenotes}
	\item [\textasteriskcentered] Unsupported functions
	\item [$\dagger$] Intentionally incorrect sensitivity to demonstrate contribution bounding
	\end{tablenotes}
	\end{small}
\end{threeparttable}
\end{table*}

We assess accuracy experimentally using TPC-H~\cite{council2008tpc}, an industry standard SQL benchmark. The TPC-H benchmarks contains a database schema and queries that are similar to those used by analysts of personal data at real-world organizations. In addition, the queries contain interesting features such as joins and a variety of aggregations. We generate a TPC-H database with the default scale factor of $1$. We treat suppliers or customers as ``users'', as appropriate.  Our metric for accuracy is median relative error, the same one used in~\cite{johnson2018towards}; a smaller median relative error corresponds to higher utility.

\subsection{Aggregation functions}

We compute the median relative error of 1,000,000 runs for our model over TPC-H Query 1 using three different aggregation functions and $\eps=0.1$ in \autoref{comparison-table}. We compare our results to 1,000,000 runs of Flex over the same query, and 10,000 runs (due to performance considerations) of PINQ over the same query. To present a fair comparison, we disabled $\tau$-thresholding and compared only one result record to remove the need for $C_u$ stability bounding. In addition, each run of the experiment was performed using a function of fixed sensitivity, controlled by supplying the function with a lower bound of $0$ and an upper bound of the value in the $\Delta_u$Q1 column. The bounds were fixed to minimize accuracy loss from contribution clamping.

For our experiments with PINQ and Flex, we also set sensitivity to our previously determined $\Delta_u$Q1 listed in \autoref{comparison-table}. The results are close to our model's results, but because neither PINQ nor Flex can \emph{enforce} contribution bounds for databases with multiple contributions per user, incorrectly set sensitivity can result in query results that are \emph{not} differentially private. Such incorrectly set bounds can be seen in experiments in Johnson et al.~\cite{johnson2018towards} and McSherry's analysis~\cite{ubernotdp}, and in the last row of \autoref{comparison-table}, where PINQ and Flex report errors far below what are required to satisfy the \eDP predicate.

With correctly set sensitivity bounds our model's results are comparable to PINQ's results for count and average. Implementation differences in our median function mean that our error is lower by a factor of 2. Both PINQ and our model outperform Flex's result for count by around an order of magnitude. We don't report errors for average and median for Flex because Flex does not support those functions.

\subsection{Aggregations with joins}

\begin{table*}
\centering
\begin{threeparttable}
    \caption{Selected TPC-H join query results ($\eps=0.1$)}
	\label{join-table}
	\begin{tabular}{llllll}
		\toprule
		Query & $\Delta_u$Q & $C_u$ &  Experimental error & $\tau$-thresholding rate & $\delta$ \\
		\midrule
		Q4 & $5$ & $5$ & $0.0339$ & $5.40\times10^{-5}$ & $6.78\times10^{-7}$\\
		Q13 & $1$ & $1$ & $0.00677$ & $0.309$ & $6.78\times10^{-7}$ \\
		Q16 & $1$ & $5$ & $11.3$\tnote{\textasteriskcentered} & $0.9999606$ & $2.07\times10^{-4}$ \\
		Q21 & $1$ & $1$ & $1.60$\tnote{\textasteriskcentered} & $0.999796$ & $2.07\times10^{-4}$ \\
		\bottomrule
	\end{tabular}
	\begin{small}
	\begin{tablenotes}
	\item [\textasteriskcentered] Results uninterpretable due to high levels of $\tau$-thresholding
	\end{tablenotes}
	\end{small}
\end{threeparttable}
\end{table*}

We present the results of running our system over a selection of TPC-H queries containing joins in Table \ref{join-table}. Similarly, we report the median relative error of 1,000,000 runs for each query using $\eps=0.1$. We report the impact of $\tau$-thresholding (the ratio of suppressed records), suggesting that our model is \edDP. $\delta$ was set with $\delta=n^{-\eps \log n}$ \cite{dwork2009differential}, where $n$ is the number of distinct users in the underlying database: either customers or suppliers, depending on the query.

Q4 represents how our system behaves when very little $\tau$-thresholding occurs. Q16 and Q21 demonstrate the opposite, both queries exhibit a very large error that differs from the theoretical error due to most partitions being removed by the threshold because of their small user count. Indeed, this is by design: as partition user counts approach 1, the ratio of $\tau$-thresholding approaches $(1-\delta)^{1/C_u}$. Finally, Q13 represents a more typical result, a query containing mostly medium user count partitions with a long tail of lower count partitions. A moderate amount of $\tau$-thresholding occurs which increases error compared to Q4, but the results are still quite accurate.

\subsection{Impact of parameters on utility}

In this section we explore the relationship between utility and various parameters, which must be adjusted to balance privacy and utility~\cite{amin2019}.

\vspace{-0.5cm}
\paragraph{Effect of $\eps$}

The privacy parameter $\eps$ is inversely proportional to the Laplace scale parameter used by anonymous functions to add noise. Hence, an increase in $\eps$ leads to a decrease in utility. The median error from noise, $x$, satisfies:
\[
x = \frac{\log(2) \Delta_u}{\eps}
\]
where $\Delta_u$ is the sensitivity (\autoref{app:theoryerror}). When a single query contains many aggregations, the privacy budget is split equally among them. In the presence of $N$ aggregations, each aggregation will satisfy $(\eps/N)$-differential privacy. Thus, utility degrades inversely with the number of aggregations.

\vspace{-0.5cm}
\paragraph{Effect of $\delta$ and $C_u$}

\begin{figure}[ht]
\centering
\includegraphics[width=0.45\textwidth]{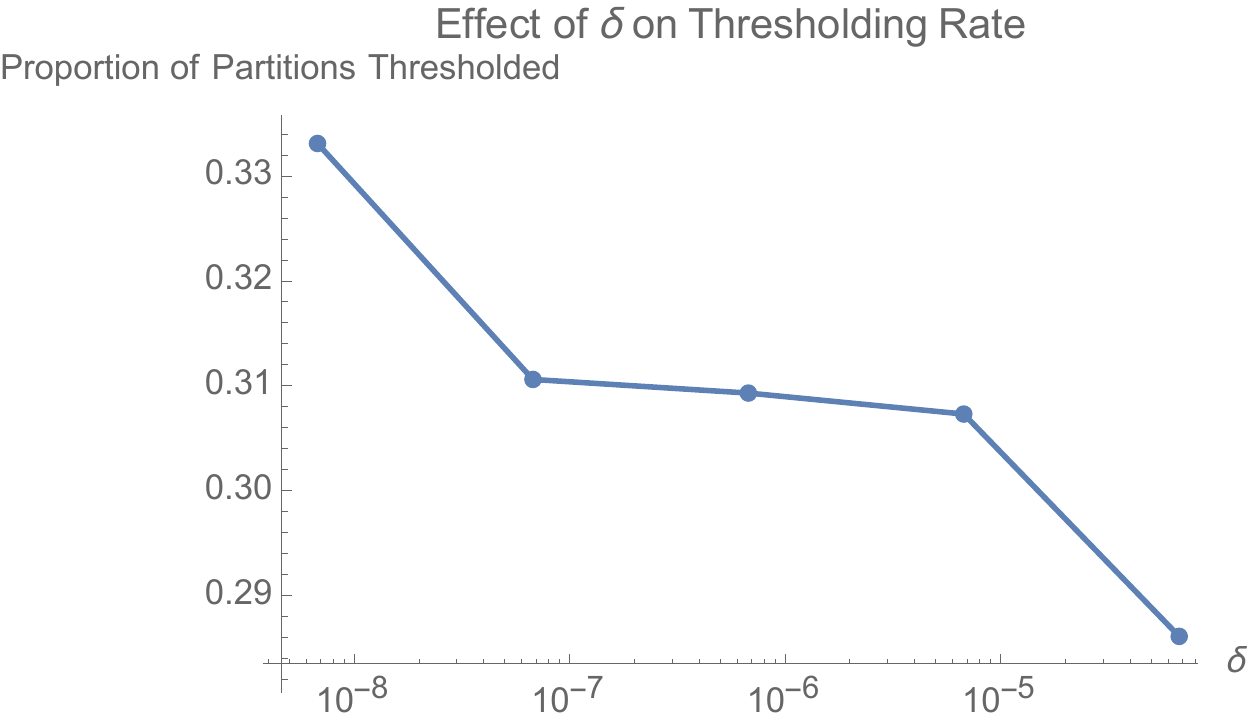}
\caption{Partition thresholding rates on Q13 induced by various $\delta$.}
\label{fig:deltaexperiment}
\end{figure}

For a fixed $\eps$, varying $\delta$ causes the threshold $\tau$ to change, which changes the number of records dropped due to thresholding. Similarly, changing $C_u$ modifies the number of records dropped due to contribution bounding. We first perform experiments on TPC-H Query 13 with $\eps=.1$ and varying $\delta$ to quantify the impact on partitions returned: \autoref{fig:deltaexperiment} displays the results. The figure shows that as $\del$ increases exponentially, the proportion of partitions thresholded decreases somewhat linearly.

Next, we analyze the effect of $C_u$ for a specific artificial query. Consider the following query \emph{after} rewriting. 

\begin{lstlisting}
SELECT ANON_COUNT(*)
FROM (SELECT uid, ROW_NUMBER() as rn
      FROM Table1
      GROUP BY uid, rn)
TABLESAMPLE RESERVOIR
    ($C_u$ ROWS PARTITION BY uid);
\end{lstlisting}

Let $U$ be the set of users in \verb|Table|, and suppose that there are $N$ users. Suppose each user $u \in U$ has a number of rows distributed according to $\mathcal{D}_u$. Then the distribution of the error in the count due to reservoir sampling is:

\[
\text{Error}_{C_u} = \sum_{u\in U} \mathcal{D}_u - ( \mathcal{D}_u | \mathcal{D}_u \leq C_u )
\]

\begin{figure}[ht]
\centering
\includegraphics[width=0.45\textwidth]{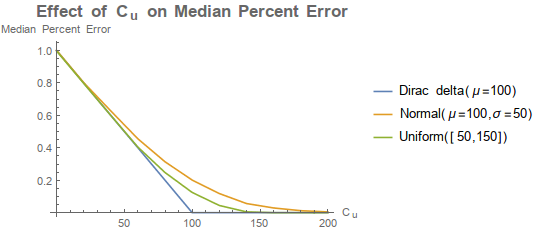}
\caption{Median percent error induced by $C_u$ for various distributions centered at 100.}
\label{fig:cuerror}
\end{figure}

We divide the median of $\text{Error}_{C_u}$ by the total expected count $\sum_{u\in U} E[\mathcal{D}_u]$ to obtain the median percent error.
\autoref{fig:cuerror} shows the effect of $C_u$ on median percent error with $N=1000$ and various distributions $\mathcal{D}_u$. All distributions have similar behavior as $C_u$ increases, but the median percent error declines at different speeds based on distribution shape.

\vspace{-0.5cm}
\paragraph{Effect of clamping}\label{effectclamping}

We analyze the effect of clamping on accuracy using model input distributions. Since clamping occurs at the \eDP aggregation level, we focus on input sets that have at most one row per user. 

Consider finding \verb|ANON_AVG(S, l, u)|, where $S$ is size $N$ and uniformly distributed on $[a, b]$. For symmetric input distribution, symmetric clamping will not create bias, so we clamp only one end: consider clamping bounds $(l, u)$ such that $l = a$ and $a < u < b$. We analyze expected error since median error is noisier when running experiments, and the behavior of both metrics are similar. 

\begin{figure}[ht]
\centering
\includegraphics[width=0.45\textwidth]{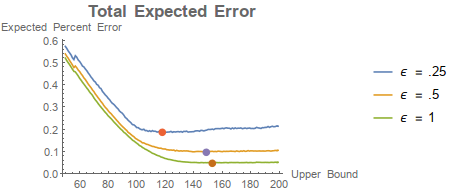}
\caption{Total median error for various $\eps$.}
\label{fig:clampingmedianerror}
\end{figure}

We plot the impact of the upper clamp bound on total expected error for uniform $S$ with $(a, b) = (50, 150)$ in \autoref{fig:clampingmedianerror}. We used lower bound $l = -200$, $N=100$, and various $\eps$. The optimal point on each curve is marked with a circle. To maximize accuracy, overestimating the spread of the input set must be balanced with restricting the sensitivity. Analysis with the other aggregation functions yields similar results. 

We perform further clamping analysis with multiple distributions in \autoref{app:clampinganalysis}.

\section{Practical considerations}

Designing a differentially private query engine for non-expert use requires a number of considerations beside the design and implementation of the system described in the previous section. In this section, we highlight a few of these concerns, and detail the approaches we have taken to mitigate them.

\subsection{Usability}

In this section, we present the ways we improved the system's usability.

\vspace{-0.5cm}
\paragraph{Automatic bounds determination}\label{approxbounds}

One major difference between standard SQL queries and queries using our differentially private aggregation operator is the presence of bounds: e.g., when using \ \verb|ANON_SUM|, an analyst must specify the lower and upper bound for each sum input. This differs from standard workflows, and more importantly, it requires prior knowledge of the data that an analyst might not have.

To remove this hurdle, we designed an aggregation function which can be sequentially composed with our previously introduced \eDP functions to automatically compute bounds that minimize accuracy loss. Call this function \verb|APPROX_BOUNDS(col)|.

Without contribution bounds, the domain of an \eDP function $f$ in our model spans $\mathbb{R}$. Finding differentially private extrema over such a range is difficult. Fortunately, we can leverage two observations. First, inputs to $f$ are represented on a machine using finite precision; typically 64-bit integers or floating point numbers. Second, bounds do not need to be close to the real extrema: clipping a small fraction of data will usually not have a large influence on the aggregated results, and might even have a positive influence by removing outliers.

Consider an \verb|ANON_SUM| operation operating on 64-bit unsigned integers where bounds are not provided. We divide the privacy budget in two: the first half to be used to infer approximate bounds using \verb|APPROX_BOUNDS(col)|; the second half to be used to calculate the noisy sum as usual. We must spend privacy budget to choose bounds in a data-dependent way.

In the \verb|APPROX_BOUNDS(col)| function, we instantiate a 64-bin logarithmic histogram of base 2, and, for each input $i$, increment the $\lceil \log_2 i \rceil$th bin. Laplace noise is then added to the count in each bin, as is standard for differentially private histograms~\cite{dwork2014algorithmic}.
Then, to find the approximate maximum of the input values, we select the most significant bin whose count exceeds some threshold $t$, calculated using parameters $B$ and $P$:
$$t = \frac{1}{\eps} \log (1 - P^{\frac{1}{B-1}})$$

where $B$ is the count of histogram bins and $P$ is the desired probability of \emph{not} selecting a false positive. For example, $B = 64$ for unsigned integers. For the derivation of the threshold $t$, see~\autoref{app:autoboundthresh}.

When setting $P$, the trade-offs of clipping distribution tails, false positive risk, and algorithm failure due to no bin count exceeding $t$ must all be considered. Values on the order of $(1-10^{-9})$ for $P$ can be suitable, depending on $\eps$ and the size of the input database.

The approximate minimum bound can similarly be found by searching for the least significant bin with count exceeding~$t$. We generalize this for signed numbers by adding additional negative-signed bins, and for floating point numbers by adding bins for negative powers of $2$.

\vspace{-0.5cm}
\paragraph{Representing accuracy and privacy}

A ubiquitous challenge for a DP interface is the fact that acceptable accuracy loss is data-dependent. We address this by giving analysts a variety of utility loss measures to make an informed decision.

For each result, we attach a confidence interval (CI) of the noise that was added. The CI can be calculated from each function's contribution bounds and share of $\eps$. The CI does not account for the effect of clamping or thresholding. In addition, during automatic bounds determination (\autoref{approxbounds}), the log-scale histogram gives us an approximate fraction of inputs exceeding the chosen bounds; this can also be returned to the analyst.

For queries with a long tail of low user count partitions that do not pass $\tau$-thresholding, we can combine all such partitions into a single partition. If the combined partition now exceeds the $\tau$-threshold, we may return aggregate results for the "leftovers" partition. This will allow analysts to estimate data loss.

To represent privacy, there are well-established techniques~\cite{lee2011much,hsu2014differential,naldi2015differential,krehbiel2019choosing} and perspectives~\cite{nissim2017differential} in the literature.

\subsection{Manual testing}

Testing is necessary to get a strong level of assurance that our query engine correctly enforces its privacy guarantee. We audited the code manually and found some implementation issues. Some of these issues have previously been explored in the literature, notably regarding the consequences of using a floating-point representation~\cite{mironov2012significance} with Laplace noise. Some of them, however, do not appear to have been previously mentioned in the literature, and are good examples of what can go wrong when writing secure anonymization implementations.

One of these comes from another detail of floating-point representation: special \verb|NaN| (``not a number'') values. These special values represent undefined numbers, like the result of $0/0$. Importantly, arithmetic operations including a \verb|NaN| are always \verb|NaN|, and comparisons between any \verb|NaN| and other numbers are always \verb|False|. This can be exploited by an attacker, for example using a query like \verb|ANON_SUM(IF uid=4217 THEN 0/0 ELSE 0)|. The \verb|NaN| value will survive naive contribution bounding (bounds checks like \verb|if(value > upper_bound)| will return \verb|False|), and the overall sum will return \verb|NaN| iff the condition was verified. We suspect that similar issues might arise with the use of special \verb|infinity| values, although we have not found them in our system (such values are correctly clamped).

From this example, we found that a larger class of issues can appear whenever the user can abuse a branching condition to fail if an arbitrary condition is satisfied (by example, by throwing a runtime error or crashing the engine). Thus, in a completely untrusted environment, the engine should catch all errors and silently ignore them, to avoid leaking information in the same way; and it should be hardened against crashes. We do not think that we can completely mitigate this problem, and silently catching all errors severely impedes usability. Thus, is it a good idea to add additional risk mitigation techniques, like query logging and auditing.

Interestingly, fixing the floating-point issue in~\cite{mironov2012significance} leads to a \emph{different} issue when using the Laplace mechanism in $\tau$-thresholding. The secure version of Laplace mechanism requires rounding the result to the nearest $r$, where $r$ is the smallest power of 2 larger than $1/\eps$. If the $\tau$-thresholding check is implemented as $\verb|if (noisy_count >= tau)|$, then a noisy count of e.g. $38.1$ can be rounded up to e.g. $40$ (with $r=4$). If the threshold $\tau$ is $39$, and the $\del$ calculation is based on a theoretical Laplace distribution, then a noisy count of $38.1$ shouldn't pass the threshold, but will: this leads to underestimating the true $\del$. This can be fixed by using a non-rounded version of the Laplace mechanism for thresholding only; as the numerical output is never displayed to the user, attacks described in~\cite{mironov2012significance} don't apply.

\subsection{Stochastic testing}\label{sec:stochastictesting} 

While the operations used in the engine are theoretically proven to be differentially private, it is crucial to verify that these operations are implemented correctly. Since the number of possible inputs is unbounded, it is impossible to exhaustively test this. Thus we fall back to \textit{stochastic testing} and try to explore the space of databases as efficiently as possible. This does not give us a guarantee that an algorithm passing the test is differentially private, but it is a good mechanism to detect violations.

Note that we focus on testing DP \emph{primitives} (aggregation functions) in isolation, which allows us to restrict the scope of the tests to row-level DP. We then use classical unit testing to independently test contribution bounding. We leave it as future work to extend our system to handle end-to-end user-level DP testing.

Our testing system contains four components: database generation, search procedure to find database pairs, output generation, and predicate verification.

\vspace{-0.5cm}
\paragraph{Database generation and testing}\label{sec:st:database-generation}

\begin{figure}[ht]
\centering
\includegraphics[width=0.4\textwidth]{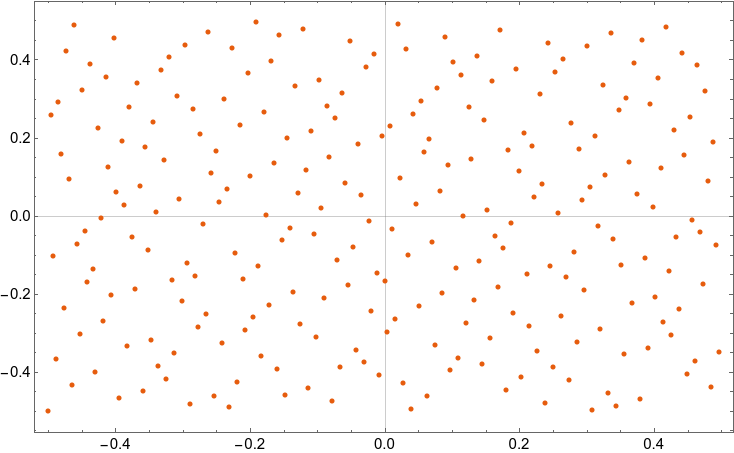}
\caption{256 points over a $[-0.5, 0.5]^2$ unit square: Halton sequence}
\label{fig:halton-sequence}
\end{figure}

What databases should we be generating? All DP aggregation functions are scale-invariant, so without loss of generality, we can consider only databases with values in a unit range $[-r, r]$. Of course, we can't enumerate all possible databases $[-r, r]^S$, where $S$ is the size of the database. Instead, we try to generate a \emph{diverse} set of databases. We use the Halton sequence~\cite{Halton1964} to do so. As an example, \autoref{fig:halton-sequence} plots databases of size 2 generated by a Halton sequence. Unlike uniform random sampling, Halton sequences ensure that databases are evently distributed and not clustered together.

\begin{figure}[ht]
\centering
\includegraphics[width=0.3\textwidth]{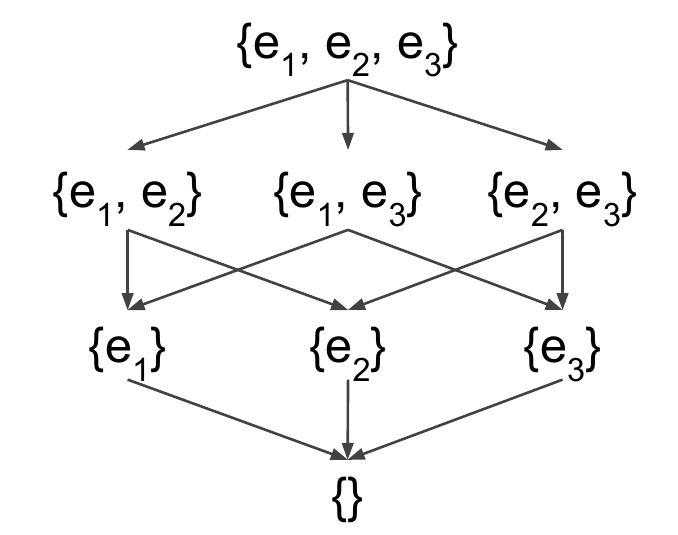}
\caption{Database Search Graph for a database $\{e_1, e_2, e_3\}$}
\label{fig:database-graph}
\end{figure}

A database is a set of records: we consider its power set, and find database pairs by recursively removing records. This procedure is shown in \autoref{fig:database-graph}.

\vspace{-0.5cm}
\paragraph{DP predicate test}

Once we have pairs of adjacent databases, we describe how we test each pair
$\left(D_1,D_2\right)$. The goal is to check that for all possible outputs $S$ of mechanism $f$:
$$\Pr[f(D_1) \in S] \leq e^{\eps} \Pr[f(D_2) \in S] + \delta.$$

By repeatedly evaluating $f$ on each database, we estimate the density of these probability distributions. We then use a simple method to compare these distributions: histograms.

\begin{figure}
\centering
\begin{subfigure}[b]{0.2\textwidth}
\centering
\includegraphics[width=\textwidth]{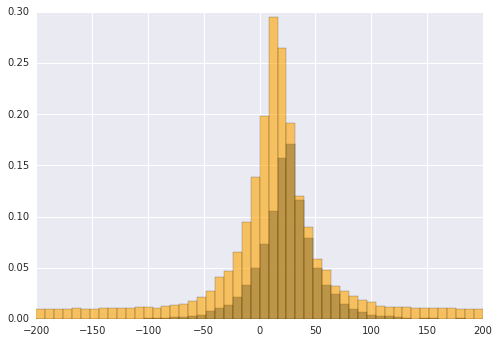}
\caption{Passing test}
\label{fig:hist-dp}
\end{subfigure}
\begin{subfigure}[b]{0.2\textwidth}
\centering
\includegraphics[width=\textwidth]{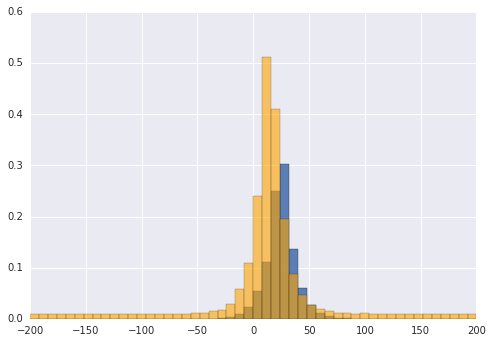}
\caption{Failing test}
\label{fig:hist-non-dp}
\end{subfigure}
\caption{Histogram examples for DP testing, given one pair of databases}
\end{figure}

We illustrate this procedure in \autoref{fig:hist-dp} and \autoref{fig:hist-non-dp}. The upper curves (in orange) are the upper DP bound, created by multiplying the probability estimate of each bin for database $D_1$ by $e^{\eps}$ and adding $\delta$. The lower curve (in blue) is the unmodified probability estimate of $D_2$. In \autoref{fig:hist-dp}, all blue buckets are less than the upper DP bound: that the DP predicate is not violated.  In \autoref{fig:hist-non-dp}, 3 buckets exceed this upper bound: the DP predicate is been violated. For symmetry, we also repeat this check with $D_1$ swapped with $D_2$.

It is sufficient to terminate once we find a single pair of databases which violate the predicate.
However, since the histogram is subject to sampling error, a correctly implemented algorithm can fail this test with non-zero probability.
To address this, we relax our test by using confidence intervals as bounds~\cite{wasserman2013all}.
We can also parameterize the tester with a parameter $\alpha$ that tolerates a percentage of failing buckets per histogram comparison.

\vspace{-0.5cm}
\paragraph{DP stochastic tester algorithm}
The overall approach is an algorithm that iterates over databases and performs a DFS on each of the database search graphs, where each edge is a DP predicate test.  See \autoref{app:stochastic-testing-algorithm} for more details.

\vspace{-0.5cm}
\paragraph{Case study: noisy average}
We were able to detect that an algorithm was implemented incorrectly, violating DP.
When we first implemented \verb|ANON_AVG|, we used the Noisy Average with Accurate Count algorithm from~\cite{li2016differential}: we used our \verb|ANON_SUM| implementation to compute the noisy sum and then divided it by the un-noised count.
Our first version of \verb|ANON_SUM| used a Laplace distribution with scale $\frac{|U-L|}{\eps}$, where $U$ and $L$ are the upper and lower clamping bounds, which is the correct bound when used as a component of \verb|ANON_AVG|.
However, this was \emph{not} correct for noisy sum in the case when adjacent databases differ by the presence of a row. We updated the scale to $\frac{\max(|U|, |L|)}{\eps}$, as maximum change in this case is the largest magnitude. This change created a regression in DP guarantee for \verb|ANON_AVG|, which was detected by the stochastic tester.

\begin{figure}
\centering
\includegraphics[width=0.4\textwidth]{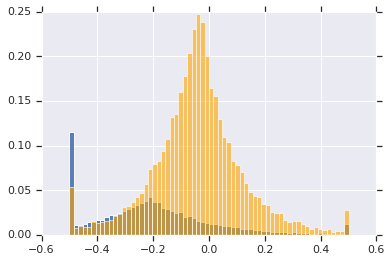}
\caption{Example histogram comparison for the Noisy Average algorithm with incorrect noise.\\
Orange: $e^{\eps} Pr[f(\lbrace -0.375, -0.055, 0.3 \rbrace)]$\\
Blue: $Pr[f(\lbrace -0.375, -0.055 \rbrace)]$}
\label{fig:noisy-average-violation}
\end{figure}

\autoref{fig:noisy-average-violation} shows a pair of datasets where the stochastic tester detected a violation of the DP predicate: $\lbrace -0.375, -0.055, 0.3 \rbrace$ and $\lbrace -0.375, -0.055 \rbrace$.
We can clearly see that several buckets violate the predicate. Once the stochastic tester alerted us to the error we quickly modified \verb|ANON_AVG| to no longer depend on \verb|ANON_SUM| so that it could use the correct sensitivity.

\section{Conclusion and future work} 

We presented a generic system to answer SQL queries with user-level differential privacy. This system is able to capture most data analysis tasks based on aggregations, performs well for typical use-cases, and provides a mechanism to deduce privacy parameters from accuracy requirements, allowing a principled decision between re-identification risk and the required utility of a particular application. All implemented mechanisms are tested with a stochastic checker that prevents regressions and increases our level of confidence in the robustness of the privacy guarantee. By releasing components of our system as open-source software after we validated its viability on internal use-cases, we hope to encourage further adoption and research of differentially private data analysis.

The algorithms presented in this work are relatively simple, but empirical evidence show that this approach is useful, robust and scalable. Future work could include usability studies to test the success of the methods we used to explain the system and its inherent accuracy/privacy trade-offs. In addition, we see room for significant accuracy improvements: using Gaussian noise and better composition theorems is a natural next step. There might also be opportunities to compensate the data loss due to contribution bounding and thresholding, optimize the algorithms used for specific sets of queries, ore use amplification by sampling. We did not attempt to cache results to allow people to re-run the queries, but further work could also explore the usability and privacy impact of such a method.

More generally, we believe that future work on DP should consider that realistic data typically includes multiple contributions for a single user: we believe that contribution bounds can be built into many other DP mechanisms that are not SQL-based.

\bibliographystyle{plain}
\bibliography{main}

\begin{thebibliography}{10}

\bibitem{amin2019}
Kareem Amin, Alex Kulesza, Andres Munoz, and Sergei Vassilvtiskii.
\newblock Bounding user contributions: A bias-variance trade-off in
  differential privacy.
\newblock In {\em Proceedings of the 36th International Conference on Machine
  Learning, PMLR 97}, pages 263--271, 2019.

\bibitem{bater2018shrinkwrap}
Johes Bater, Xi~He, William Ehrich, Ashwin Machanavajjhala, and Jennie Rogers.
\newblock Shrinkwrap: Differentially-private query processing in private data
  federations.
\newblock {\em arXiv preprint arXiv:1810.01816}, 2018.

\bibitem{ben2008bayesian}
Michael Ben-Or and Avinatan Hassidim.
\newblock The {Bayesian} learner is optimal for noisy binary search (and pretty
  good for quantum as well).
\newblock In {\em 2008 49th Annual IEEE Symposium on Foundations of Computer
  Science}, pages 221--230. IEEE, 2008.

\bibitem{bichsel2018dp}
Benjamin Bichsel, Timon Gehr, Dana Drachsler-Cohen, Petar Tsankov, and Martin
  Vechev.
\newblock {DP}-finder: Finding differential privacy violations by sampling and
  optimization.
\newblock In {\em Proceedings of the 2018 ACM SIGSAC Conference on Computer and
  Communications Security}, pages 508--524. ACM, 2018.

\bibitem{bindschaedler2017plausible}
Vincent Bindschaedler, Reza Shokri, and Carl~A Gunter.
\newblock Plausible deniability for privacy-preserving data synthesis.
\newblock {\em Proceedings of the VLDB Endowment}, 10(5):481--492, 2017.

\bibitem{bun2016concentrated}
Mark Bun and Thomas Steinke.
\newblock Concentrated differential privacy: Simplifications, extensions, and
  lower bounds.
\newblock In {\em Theory of Cryptography Conference}, pages 635--658. Springer,
  2016.

\bibitem{council2008tpc}
Transaction Processing~Performance Council.
\newblock {TPC-H} benchmark specification.
\newblock \url{http://www.tpc.org/tpch/}, 2008.

\bibitem{desfontaines2019sok}
Damien Desfontaines and Bal{\'a}zs Pej{\'o}.
\newblock Sok: Differential privacies.
\newblock {\em arXiv preprint arXiv:1906.01337}, 2019.

\bibitem{DingEtAl2018}
Zeyu Ding, Yuxin Wang, Guanhong Wang, Danfeng Zhang, and Daniel Kifer.
\newblock Detecting violations of differential privacy.
\newblock In {\em Proceedings of the 2018 ACM SIGSAC Conference on Computer and
  Communications Security}, CCS '18, pages 475--489, New York, NY, USA, 2018.
  ACM.

\bibitem{dwork2007ad}
Cynthia Dwork.
\newblock An ad omnia approach to defining and achieving private data analysis.
\newblock In {\em International Workshop on Privacy, Security, and Trust in
  KDD}, pages 1--13. Springer, 2007.

\bibitem{dwork2009differential}
Cynthia Dwork.
\newblock The differential privacy frontier.
\newblock In {\em Theory of Cryptography Conference}, pages 496--502. Springer,
  2009.

\bibitem{dwork2006calibrating}
Cynthia Dwork, Frank McSherry, Kobbi Nissim, and Adam Smith.
\newblock Calibrating noise to sensitivity in private data analysis.
\newblock In {\em Theory of Cryptography Conference}, pages 265--284. Springer,
  2006.

\bibitem{dwork2010pan}
Cynthia Dwork, Moni Naor, Toniann Pitassi, Guy~N Rothblum, and Sergey Yekhanin.
\newblock Pan-private streaming algorithms.
\newblock In {\em ICS}, pages 66--80, 2010.

\bibitem{dwork2014algorithmic}
Cynthia Dwork and Aaron Roth.
\newblock The algorithmic foundations of differential privacy.
\newblock {\em Foundations and Trends in Theoretical Computer Science},
  9(3--4):211--407, 2014.

\bibitem{francis2017diffix}
Paul Francis, Sebastian~Probst Eide, and Reinhard Munz.
\newblock Diffix: High-utility database anonymization.
\newblock In {\em Annual Privacy Forum}, pages 141--158. Springer, 2017.

\bibitem{geng2012optimal}
Quan Geng and Pramod Viswanath.
\newblock The optimal mechanism in differential privacy.
\newblock {\em arXiv preprint arXiv:1212.1186}, 2012.

\bibitem{gotz2011publishing}
Michaela Gotz, Ashwin Machanavajjhala, Guozhang Wang, Xiaokui Xiao, and
  Johannes Gehrke.
\newblock Publishing search logs—a comparative study of privacy guarantees.
\newblock {\em IEEE Transactions on Knowledge and Data Engineering},
  24(3):520--532, 2011.

\bibitem{Halton1964}
J.~H. Halton.
\newblock Algorithm 247: Radical-inverse quasi-random point sequence.
\newblock {\em Commun. ACM}, 7(12):701--702, December 1964.

\bibitem{hsu2014differential}
Justin Hsu, Marco Gaboardi, Andreas Haeberlen, Sanjeev Khanna, Arjun Narayan,
  Benjamin~C Pierce, and Aaron Roth.
\newblock Differential privacy: An economic method for choosing epsilon.
\newblock In {\em 2014 IEEE 27th Computer Security Foundations Symposium},
  pages 398--410. IEEE, 2014.

\bibitem{flexgithub}
Noah Johnson and Joseph~P Near.
\newblock Dataflow analysis \& differential privacy for {SQL} queries.
\newblock \url{https://github.com/uber/sql-differential-privacy}.
\newblock Accessed: 2019-09-04.

\bibitem{johnson2018towards}
Noah Johnson, Joseph~P Near, and Dawn Song.
\newblock Towards practical differential privacy for {SQL} queries.
\newblock {\em Proceedings of the VLDB Endowment}, 11(5):526--539, 2018.

\bibitem{kairouz2017composition}
Peter Kairouz, Sewoong Oh, and Pramod Viswanath.
\newblock The composition theorem for differential privacy.
\newblock {\em IEEE Transactions on Information Theory}, 63(6):4037--4049,
  2017.

\bibitem{karp2007noisy}
Richard~M Karp and Robert Kleinberg.
\newblock Noisy binary search and its applications.
\newblock In {\em Proceedings of the eighteenth annual ACM-SIAM symposium on
  Discrete algorithms}, pages 881--890. Society for Industrial and Applied
  Mathematics, 2007.

\bibitem{kifer2011no}
Daniel Kifer and Ashwin Machanavajjhala.
\newblock No free lunch in data privacy.
\newblock In {\em Proceedings of the 2011 ACM SIGMOD International Conference
  on Management of data}, pages 193--204. ACM, 2011.

\bibitem{korolova2009releasing}
Aleksandra Korolova, Krishnaram Kenthapadi, Nina Mishra, and Alexandros
  Ntoulas.
\newblock Releasing search queries and clicks privately.
\newblock In {\em Proceedings of the 18th international conference on World
  wide web}, pages 171--180. ACM, 2009.

\bibitem{kotsogiannis2019privatesql}
Ios Kotsogiannis, Yuchao Tao, Xi~He, Maryam Fanaeepour, Ashwin Machanavajjhala,
  Michael Hay, and Gerome Miklau.
\newblock Privatesql: a differentially private sql query engine.
\newblock {\em Proceedings of the VLDB Endowment}, 12(11):1371--1384, 2019.

\bibitem{kotsogiannisarchitecting}
Ios Kotsogiannis, Yuchao Tao, Ashwin Machanavajjhala, Gerome Miklau, and
  Michael Hay.
\newblock Architecting a differentially private {SQL} engine.
\newblock In {\em Conference on Innovative Data Systems Research}, 2019.

\bibitem{krehbiel2019choosing}
Sara Krehbiel.
\newblock Choosing epsilon for privacy as a service.
\newblock {\em Proceedings on Privacy Enhancing Technologies},
  2019(1):192--205, 2019.

\bibitem{lee2011much}
Jaewoo Lee and Chris Clifton.
\newblock How much is enough? choosing $\varepsilon$ for differential privacy.
\newblock In {\em International Conference on Information Security}, pages
  325--340. Springer, 2011.

\bibitem{li2014data}
Chao Li, Michael Hay, Gerome Miklau, and Yue Wang.
\newblock A data-and workload-aware algorithm for range queries under
  differential privacy.
\newblock {\em Proceedings of the VLDB Endowment}, 7(5):341--352, 2014.

\bibitem{li2016differential}
Ninghui Li, Min Lyu, Dong Su, and Weining Yang.
\newblock Differential privacy: From theory to practice.
\newblock {\em Synthesis Lectures on Information Security, Privacy, \& Trust},
  8(4):1--138, 2016.

\bibitem{mcsherrydatasynthesis}
Frank~D McSherry.
\newblock Synthethic data via differential privacy.
\newblock
  \url{https://github.com/frankmcsherry/blog/blob/master/assets/Synth-SIGMOD.pdf}.
\newblock Accessed: 2019-05-28.

\bibitem{ubernotdp}
Frank~D McSherry.
\newblock Uber's differential privacy .. probably isn't.
\newblock
  \url{https://github.com/frankmcsherry/blog/blob/master/posts/2018-02-25.md}.
\newblock Accessed: 2019-03-22.

\bibitem{mcsherry2009privacy}
Frank~D McSherry.
\newblock Privacy integrated queries: an extensible platform for
  privacy-preserving data analysis.
\newblock In {\em Proceedings of the 2009 ACM SIGMOD International Conference
  on Management of data}, pages 19--30. ACM, 2009.

\bibitem{meiser2018tight}
Sebastian Meiser and Esfandiar Mohammadi.
\newblock Tight on budget?: Tight bounds for $r$-fold approximate differential
  privacy.
\newblock In {\em Proceedings of the 2018 ACM SIGSAC Conference on Computer and
  Communications Security}, pages 247--264. ACM, 2018.

\bibitem{mironov2012significance}
Ilya Mironov.
\newblock On significance of the least significant bits for differential
  privacy.
\newblock In {\em Proceedings of the 2012 ACM conference on Computer and
  communications security}, pages 650--661. ACM, 2012.

\bibitem{mironov2017renyi}
Ilya Mironov.
\newblock R{\'e}nyi differential privacy.
\newblock In {\em 2017 IEEE 30th Computer Security Foundations Symposium
  (CSF)}, pages 263--275. IEEE, 2017.

\bibitem{naldi2015differential}
Maurizio Naldi and Giuseppe D'Acquisto.
\newblock Differential privacy: an estimation theory-based method for choosing
  epsilon.
\newblock {\em arXiv preprint arXiv:1510.00917}, 2015.

\bibitem{narayan2012djoin}
Arjun Narayan and Andreas Haeberlen.
\newblock {DJoin}: differentially private join queries over distributed
  databases.
\newblock In {\em Presented as part of the 10th $\{$USENIX$\}$ Symposium on
  Operating Systems Design and Implementation ($\{$OSDI$\}$ 12)}, pages
  149--162, 2012.

\bibitem{nissim2007smooth}
Kobbi Nissim, Sofya Raskhodnikova, and Adam Smith.
\newblock Smooth sensitivity and sampling in private data analysis.
\newblock In {\em Proceedings of the thirty-ninth annual ACM symposium on
  Theory of computing}, pages 75--84. ACM, 2007.

\bibitem{nissim2017differential}
Kobbi Nissim, Thomas Steinke, Alexandra Wood, Micah Altman, Aaron Bembenek,
  Mark Bun, Marco Gaboardi, David~R O’Brien, and Salil Vadhan.
\newblock Differential privacy: A primer for a non-technical audience.
\newblock In {\em Privacy Law Scholars Conf}, 2017.

\bibitem{wasserman2013all}
Larry Wasserman.
\newblock {\em All of statistics: a concise course in statistical inference}.
\newblock Springer Science \& Business Media, 2013.

\end{thebibliography}

\appendix

\section{Stochastic tester algorithm}\label{app:stochastic-testing-algorithm}

We present our algorithm putting all the pieces together in \autoref{alg:dp-tester}. For simplicity, we abstract away the generation of databases by including it as an input parameter here, which can be assumed to be generated by the Halton sequence as we described in \autoref{sec:st:database-generation}. We also do not include the confidence intervals or $\alpha$ parameter described earlier for dealing with the approximation errors. It is also possible to adaptively choose a histogram bin width, but we put an input parameter $K$ here. The general idea is a depth-first search procedure that iterates over edges of the database search graph.

\begin{algorithm}
\LinesNumbered
\caption{DP Stochastic Test}
\label{alg:dp-tester}
\KwIn{A random mechanism $f$, privacy parameters $\epsdel$, databases $\mathbf{D}$, number of samples $N$, number of histogram buckets $K$}
\KwOut{Decision on whether $f$ is differentially private}
\ForEach{$D_r \in \mathbf{D}$}{
  $S \leftarrow \{\text{root node } D_r$\} \tcp{Initialize search stack}
  \While{$S \neq \emptyset$}{
    $A \leftarrow pop(S)$\\
    $S \leftarrow S \cup \{succ(A)\}$\\
    \ForEach{$B \in succ(A)$}{
      \tcp{Generate samples}
      $X_A \leftarrow \{x_A^{(i)} \sim f(A) \mid i=1,\ldots,N\}$\\
      $X_B \leftarrow \{x_B^{(i)} \sim f(B) \mid i=1,\ldots,N\}$\\
      \tcp{Determine histogram buckets}
      $H_{\min}, H_{\max} \leftarrow \min(X_A \cup X_B), \max(X_A \cup X_B)$\\
      $h \leftarrow \frac{H_{\max} - H_{\min}}{K}$\\
      $\mathbf{B} \leftarrow \{B_k = [H_{\min} + (k-1) \cdot h, H_{\min} + k \cdot h]$\\ $\quad \mid k=1,\ldots,K\}$\\
      \ForEach{$B_k \in \mathbf{B}$}{
        \tcp{Check DP condition using approximate densities over $B_k$}
        \If{$\frac{1}{N}\sum_i^N \textbf{1}(x_A^{(i)} \in B_k) >$\\ \quad $e^{\eps} \frac{1}{N} \sum_i^N \textbf{1}(x_B^{(i) \in B_k)} + \delta$}{
          \Return $f$ is not differentially private
        }
      }
    }
  }
}
\Return $f$ is differentially private
\end{algorithm}

Our actual implementation includes all of the above omissions, including an efficient implementation of the search procedure that caches samples
of databases already generated.

\section{Proof of~\autoref{sensitivitytheorem}}\label{app:sensitivityproof}

\textbf{Restatement of \autoref{sensitivitytheorem}}.
\textit{Consider an anonymization query in the form $h(R)$, where $R$ is the input table and $h$ is the query transformation. Then there exist constants $M$, which depends on $h$, and an engine-defined constant $C_u$, such that the user sensitivity satisfies  $\Delta_u h \leq C_u M$.}

\begin{proof} Recall \autoref{def:sensitivity}:
\[
\Delta_u f = \underset{D_1,D_2\in\dbs:{||D_1-D_2||}_u=1}\max{{||f(D_1)-f(D_2)||}_1}
\]

Let $h$ be some allowed query transformation in our privacy model. Since $h$ must return a vector of aggregate values as the output, we can write $h = F \circ f$, where $F\colon \dbs \rightarrow \dbs$ and $f\colon \dbs\rightarrow \mathbb{R}^d$. In other words, $F$ is a database-to-database transformation while $f$ takes a database and returns a vector of real numbers. Suppose that $F$ has stability $c$. Then for any 
$D_1, D_2 \in \dbs$ such that $||D_1-D_2||=k$, we have  $||F(D_1) - F(D_2)|| \leq kc$. 

Suppose the maximum number of rows any user owns in the database is $k$. Then for our query $h$, the addition or deletion of a single user from a database $D_1$ creates at most $k \cdot c$ changes in $F(D_1)$. Thus, our user-global sensitivity is bounded by:
\[
\Delta_u h \leq \underset{D_1,D_2\in\dbs: ||F(D_1) - F(D_2)|| = kc}\max \; ||f(D_1)-f(D_2)||_1
\]

The databases $F(D_1)$ for all $D_1 \in \dbs$ is a subset of all databases $\dbs$, so 
\[\Delta_u h \leq  \underset{D_1,D_2\in\dbs: ||D_1-D_2|| = kc}\max \; ||f(D_1)-f(D_2)||_1,
\]
which, by the definition of global sensitivity, can be bounded as 
\[
\Delta_u h \leq kc \;\Delta h.
\]
Now, consider in addition that:
\[
\Delta h =  \underset{D_1,D_2\in\dbs: ||D_1-D_2|| = 1}\max \; ||f(F(D_1))-f(F(D_2))||_1.
\]
Again, since $F(D_1)$ for all $D_1 \in \dbs$ is a subset of $\dbs$, 
\[
\Delta h \leq \underset{D_1,D_2\in\dbs: ||D_1-D_2|| = 1}\max \; ||f(D_1)-f(D_2)||_1 = \Delta f,
\]
from which we conclude that 
\[
\Delta_u h \leq k c \; \Delta f.
\]

Our per-user sensitivity is unbounded if at least one of $k$, $c$, or $\Delta f$ are unbounded. Our privacy model, however, is formulated so that we can bound the product. 

Since $h$ is an allowed query for our privacy model, we know that $f=(f_1,\dots,f_i)$ must be a finite vector of bounded-contribution aggregation functions, as discussed in \autoref{sec:boundedagg}. Therefore, for each $i$, the per-row global sensitivity of $f_i$ is bounded and listed in \autoref{table:agg}. Each sensitivity is function-dependent, so call it $M_i$. The sensitivity of $f$ is then bounded by the sum of these sensitivities $M = \sum M_i$.

The operator $\mathcal{T}$ bounds stability by sampling a fixed number of rows per user after the per-user aggregation stage. Call this number $C_u$. Then the number of rows owned by a user in the transformed database, previously $k \cdot c$, is now bounded by $C_u$.

Putting it all together, for each user there can only be $C_u$ contributing rows to $f$, each with a bounded contribution of $M$, as determined by the analyst-specified clamp bounds in \autoref{table:agg}. We can conclude that for model-defined constants $C_u$ and $M$, we have
\[
\Delta_u h \leq C_u  M,
\]
concluding the argument.
\end{proof}

\section{Proof of~\autoref{deltatheorem}}\label{app:deltaproof}

\begin{lemma}\label{noisycountcdf}
Consider a database $D$ containing one row. The probability that an \eDP noisy count of the number of rows in $D$ will yield at least $\tau$ for any $\tau \geq 1$ is
\[
\rho_\tau = \frac{1}{2} e^{-(\tau-1)\eps} 
\]
\end{lemma}
\begin{proof}
This is a special case of Section~5.2 in~\cite{korolova2009releasing}: the noisy count is distributed as the Laplace distribution centered at the true count of $1$ with scale parameter $1/\eps$. Evaluating the CDF at $\tau$ yields $\rho_\tau$.
\end{proof}

\textbf{Restatement of Theorem 2.}
\textit{
  Let $\eps, \delta, C_u > 0$ be privacy parameters. Consider a SQL engine that, for each non-empty group in a query's grouping list, computes and releases an $\eps/C_u$-DP noisy count of the number of contributing users. For empty groups, nothing is released. A user may not influence more than $C_u$ such counts. Each count must be $\tau$ or greater in order to be released. We may set
    \[
    \tau = 1 - \frac{C_u \log(2-2(1-\delta)^{1/C_u})}{\eps}
    \]
  to provide user-level $(\eps, \del)$-DP in such an engine.
}

\begin{proof}
Let $f$ be the SQL engine operator. Consider any pair of databases $D_1, D_2$ such that $||D_1 - D_2||_u = 1$ and such that the set of non-empty groups from $D_1$ is the same as that for $D_2$; call this set $G$. The $\eps/C_u$-DP noisy count will get invoked for all groups in $G$ for both databases, and the $\tau$-thresholding applied. For all groups not in $G$, no row will be released for both databases. A change in the user may affect a maximum of $C_u$ output counts, each of which is $\eps / C_u$-DP, so by differential privacy composition rules~\cite{kairouz2017composition}, 
\[
\Pr[f(D_1) \in S] \leq e^{\eps} \Pr[f(D_2) \in S] 
\]

Next, consider empty database $D_3$. Let $U$ be the set of all outputs and let $E$ be the output set containing only the output ``no result rows are produced''. Then we have $\Pr[f(D_3) \in U\setminus E] = 0$. It remains to show that for database $D_4$ containing a single user, $\Pr[f(D_4) \in U\setminus E] \leq \del$. The $\eps$-DP count is computed by counting the number of rows, and then adding Laplace noise with scale $C_u/\eps$. Database $D_4$ contains values for a maximum of $C_u$ groups; it is only possible to produce an output row for those groups. For each groups, the probability that the noisy count will be at least $\tau$ is $\rho_\tau = \frac{1}{2} e^{-\frac{(\tau-1)\eps}{C_u}}$, by Lemma~\ref{noisycountcdf}. Then:
\[ 
\Pr[f(D_4) \in E] = 1-(\rho_\tau)^{C_u} 
\]
so we need to satisfy:
\[
\Pr[f(D_4) \in U \setminus E] = (\rho_\tau)^{C_u}   \leq \del.
\]
Solving for $\tau$ in the expression $(\rho_\tau)^{C_u}  \leq \del$:
\[
\tau \geq 1 - \frac{C_u\log(2-2(1-\delta)^{1/C_u})}{\eps}.
\]

Lastly, consider SQL engine operator $f$ and databases $D_5, D_6$ such that $D_6$ is $D_5$ with the addition of a single user. It remains to consider the case where $D_5$ and $D_6$ do not have the same set of non-empty groups. Since they differ by one user, $D_6$ may have a maximum of $C_u$ additional non-empty groups, each containing $1$ unique user; call this set of groups $G$. Call the set of the remaining groups $G'$.  Split the rows of $D_5$ into two databases: the rows that correspond to groups in $G$ and $G'$, respectively. Call the rows of $D_6$ that correspond to groups $G$ as $D_6^0$ and the rows that correspond to groups $G'$ as $D_6^1$. Note that $D_5$ only contains rows corresponding to groups $G'$. Now, split the operator $f$ into two operators $f_0$ and $f_1$: $f_0$ is $f$ with an added filter that only outputs result rows corresponding groups in $G$; $f_1$ is the same for $G'$. 

We have decomposed our problem into the previous two cases. The system of $f_0$, $D_5$, and $D_6^0$ is the case where the pair of databases have the same set of non-empty groups, $G'$. The system of $f_1$, the empty database, and $D_6^1$ is the case where the non-empty database contains a single user. Each system satisfies the DP predicate separately. Since they operate on a partition of all groups in $D_5$ and $D_6$, the two systems satisfy the DP predicate when recombined into $f$, $D_5$, and $D_6$.

We have shown that for two databases differing by a single user, the DP predicate is satisfied. Thus, we have shown that our engine provides user-level $(\eps, \del)$-DP.
\end{proof}

\section{Laplace median error}\label{app:theoryerror}

We find the theoretical median noise of a Laplace distribution. Divide the theoretical median noise by the exact result to obtain the theoretical median error. 

For instance, the \verb|ANON_COUNT| function applies Laplacian noise with parameter $\Delta_u$ $/ \varepsilon$. Let $x_{\text{count}}$ be the theoretical median noise. Then $x_{\text{count}}$ satisfies:

\begin{align*}
\frac{1}{4} &= \text{CDF}_{\text{Laplace}}(x_{\text{count}}) - \text{CDF}_{\text{Laplace}}(0) \\
&= (1 - \frac{1}{2} e^{\frac{-x_{\text{count}} \varepsilon}{\Delta_u }}) - \frac{1}{2}
\end{align*}

And thus:
\begin{align*}
x_{\text{count}} &= \frac{\log(2) \Delta_u }{\varepsilon}.
\end{align*}

\section{Automatic bounding threshold}\label{app:autoboundthresh}

In this section we will derive the internal threshold used in the \verb|APPROX_BOUNDS(col)| function described in~\autoref{approxbounds}.

We argue that for privacy parameter $\eps$, the number of histogram bins $B$, and the probability of a false positive $P$, we should set the threshold $t$ to be

\[
 t = \frac{1}{\eps} \log (1 - P^{\frac{1}{B-1}})
\]

Recall that in the automatic bounding algorithm, we create a logarithmic histogram of input values, and apply Laplace noise to the count in each histogram bin. The probability that a given bin produced a count of $x$ if its true count is zero is

\[
P_{\text{bin}} = e^{-x\eps}
\]

Suppose we are looking for the most significant bin with a count greater than $t$. In the \verb|APPROX_BOUNDS(col)| function, we iterate through the histogram bins, from most to least significant, until we find one exceeding $t$. In the worst case, the desired bin is the \emph{least} significant bin. This means $B-1$ bins with exact counts of $0$ must not have noisy counts exceeding $t$. Thus, the probability that there was not a false positive in this worst case is

\[
P = (1- e^{-t\eps} )^{B-1}
\]

Solving for $t$, we obtain the desired threshold.

\section{Aggregation sensitivity bounds}\label{app:sensitivitybound}

The definition of a bounded-sensitivity aggregate function is given in~\autoref{sec:boundedagg}. We will show that the sensitivity bounds listed in \autoref{table:agg} are valid. Some of the bounds are very loose. Note that in each of these, we only consider cases where the two databases compared in the definition of DP have one or more rows: the case we compare a empty database with a database having only one row is tackled by the $\tau$-thresholding, detailed in section~\autoref{sec:kthreshold}.

\begin{lemma}
\verb|ANON_COUNT(col)| is bounded by sensitivity 1.
\begin{proof}
Adding any row only changes the count by 1.
\end{proof}
\end{lemma}

\begin{lemma}
\verb|ANON_SUM(col, L, U)| is bounded by sensitivity $\max(|L|, |U|)$.
\begin{proof}
Consider adding a clamped input $x$. Since $L \leq x \leq U$, we have $|x| \leq |L|$ and $|x| \leq |U|$.
\end{proof}
\end{lemma}

\begin{lemma}\label{anonavgbound}
\verb|ANON_AVG(col, L, U)| is bounded by sensitivity $|U-L|$.
\begin{proof}
The average of a clamped set of inputs whose elements are on $[L, U]$ will always lie on $[L, U]$. Then the change in the average when adding or removing an element must be bounded by $|U-L|$.
\end{proof}
\end{lemma}

\begin{lemma}\label{lemma:anonvarbound}
\verb|ANON_VAR(col, L, U)| is bounded by sensitivity $ |U-L|^2$.
\begin{proof}
Consider clamped input set $Z$ of size $N$ with average $\mu$. The following is the variance.
\[
\frac{\sum_{z \in Z} (z-\mu)^2 }{N}
\]

The values $z, \mu \in [L, U]$, so the magnitude difference between them must be bounded by $|U-L|$. Then the variance is bounded by 
\[
\frac{\sum_{z \in Z} |U-L|^2 }{N} = |U-L|^2
\]
\end{proof}
\end{lemma}

\begin{lemma}
\verb|ANON_STDDEV(col, L, U)| is bounded by sensitivity $ |U-L|^2$.
\begin{proof}
This follows directly from taking the square root of the bound in the proof of \autoref{lemma:anonvarbound}.
\end{proof}
\end{lemma}

\begin{lemma}
\verb|ANON_NTILE(col, L, U)| is bounded by sensitivity $|U-L|$.
\begin{proof}
The search space is bounded by $[L, U]$ so the result is in the interval. Thus, sensitivity can never exceed the interval width.
\end{proof}
\end{lemma}

\section{Clamping analysis}\label{app:clampinganalysis}

\begin{figure}[ht]
\centering
\includegraphics[width=0.45\textwidth]{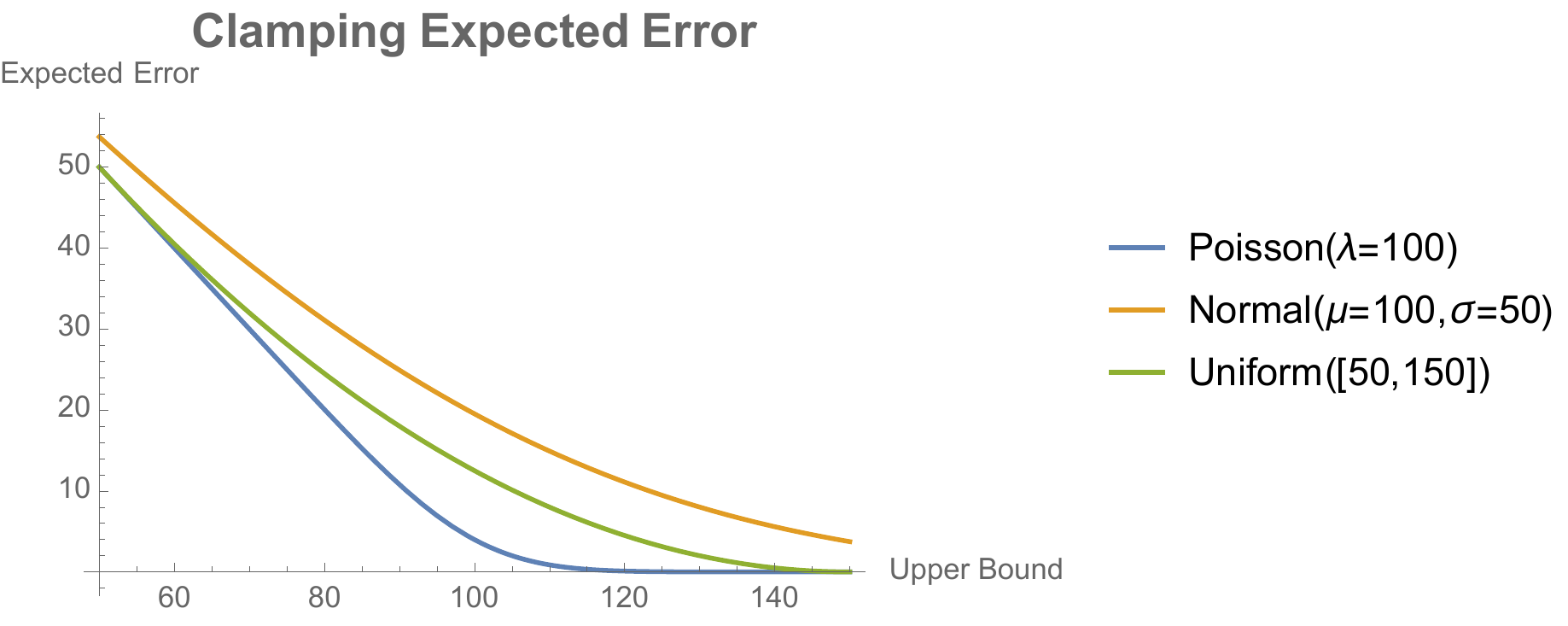}
\caption{Clamping error for distributions centered at $100$.}
\label{fig:clampingerror}
\end{figure}

Consider the problem described in \autoref{effectclamping}. The unclamped expected mean of $S$ is $\frac{a+b}{2}$.; for $S$ clamped between $[l, u]$, it is 
\[
\frac{(2au-a^2-u^2)}{2(b-a)}
\]
Thus, the expected error of the clamped mean is
\[
\frac{(b-u)^2}{2(b-a)}.
\]
Compare the expected error to the median noise added by \verb|ANON_AVG(S, l, u)|, which is approximately $\frac{\log(2)(u-l)}{N \eps}$ (\autoref{app:theoryerror}). In particular, the clamping error grows quadratically with $(b-u)$ while the noise error only grows linearly with $|u-l|$. Behavior for other distributions is similar: \autoref{fig:clampingerror} displays the clamping expected error as a function $u$ for clamping bounds $(l, u)$.

\end{document}